\RequirePackage{fix-cm}
\documentclass[smallextended]{svjour3}       
\smartqed  
\usepackage{graphicx}
\usepackage{comment}
\usepackage{hyperref}
\usepackage{array}
\usepackage{booktabs}
\usepackage{cancel}
\usepackage{mathptmx}      
\usepackage[numbers,sort&compress]{natbib}
\usepackage{hyperref}
\usepackage{amsmath,amssymb}
\usepackage{hyphenat}
\usepackage[table]{xcolor}
\usepackage{soul}

\newcommand{\kB}{k_{\rm{B}}}

\newcommand{\trm}[1]{{\rm{#1}}}
\newcommand{\rh}{\mathrm{RH}}

\definecolor{umb}{rgb}{0.25, 0.4, 0.96}
\newcommand{\rev}[1]{{\textcolor{black}{#1}}}

\def\rmd{{\mathrm{d}}}

\def\rme{{\mathrm{e}}}

\def\Equation{\textcolor{blue}{Equation}}
\def\Eq{\textcolor{blue}{Eq.}}
\def\Eqs{\textcolor{blue}{Eqs.}}
\def\Figure{\textcolor{blue}{Figure}}
\def\Fig{\textcolor{blue}{Fig.}}
\def\Sec{\textcolor{blue}{Sec.}}

\def\Tab{\textcolor{blue}{Table}}

\def\ie{{\em i.e.}}
\def\eg{{\em e.g.}}
\def\ea{{\em et al.}}
\journalname{Journal of Biological Physics}
\begin{document}

\title{Relative humidity in droplet and airborne transmission of disease}


\author{An\v{z}e Bo\v{z}i\v{c}         \and
        Matej Kandu\v{c}
}

\institute{A. Bo\v zi\v c (corr. author)\at
              Dept. of Theoretical Physics, Jo\v zef Stefan Institute, Ljubljana, Slovenia\\
              \email{anze.bozic@ijs.si}
           \and
           M. Kandu\v c (corr. author)\at
              Dept. of Theoretical Physics, Jo\v zef Stefan Institute, Ljubljana, Slovenia \\
              \email{matej.kanduc@ijs.si}
}

\date{Received: date / Accepted: date}

\maketitle

\begin{abstract}
A large number of infectious diseases is transmitted by respiratory droplets. How long these droplets persist in the air, how far they can travel, and how long the pathogens they might carry survive are all decisive factors for the spread of droplet-borne diseases. The subject is extremely multifaceted and its aspects range across different disciplines, yet most of them have only seldom been considered in the physics community. In this review, we discuss the physical principles that govern the fate of respiratory droplets and any viruses trapped inside them, with a focus on the role of relative humidity. Importantly, low relative humidity---as encountered, for instance, indoors during winter and inside aircraft---facilitates evaporation and keeps even initially large droplets suspended in air as aerosol for extended periods of time. What is more, relative humidity affects the stability of viruses in aerosol through several physical mechanisms such as efflorescence and inactivation at the air-water interface, whose role in virus inactivation nonetheless remains poorly understood. Elucidating the role of relative humidity in the droplet spread of disease would permit us to design preventive measures that could aid in reducing the chance of transmission, particularly in indoor environment.
\end{abstract}

\section{Introduction}
\label{sec:intro}

One of the prevalent ways in which numerous viruses, bacteria, and fungi spread among plants, animals, and humans is by droplets of various sizes~\cite{Fernstrom2013,laRosa2013,Verreault2008}. Humans produce respiratory droplets during talking, coughing, sneezing, and other similar activities~\cite{Belser2010,Gralton2011,Nazaroff2016,Tellier2009,Thomas2013}. These droplets, which can potentially contain pathogens, then spread outside the human body in different ways, enabling the pathogens to find a new host~\cite{Fernstrom2013,Kutter2018,laRosa2013,Verreault2008,Huang2020covid}. Most of the droplets deposit on various objects (\eg, buttons, door knobs, tabletops, and touchscreens), turning them into infectious ``fomites''. The droplets can also be inhaled by another person in close proximity ($\approx1$ to $2$~m), which provides a direct path for infection. And not least, some (particularly small) droplets can remain airborne for longer periods of time and travel considerable distances, providing yet another important path for disease transmission.

\begin{figure}[!b]
\centering
\includegraphics[width=0.95\linewidth]{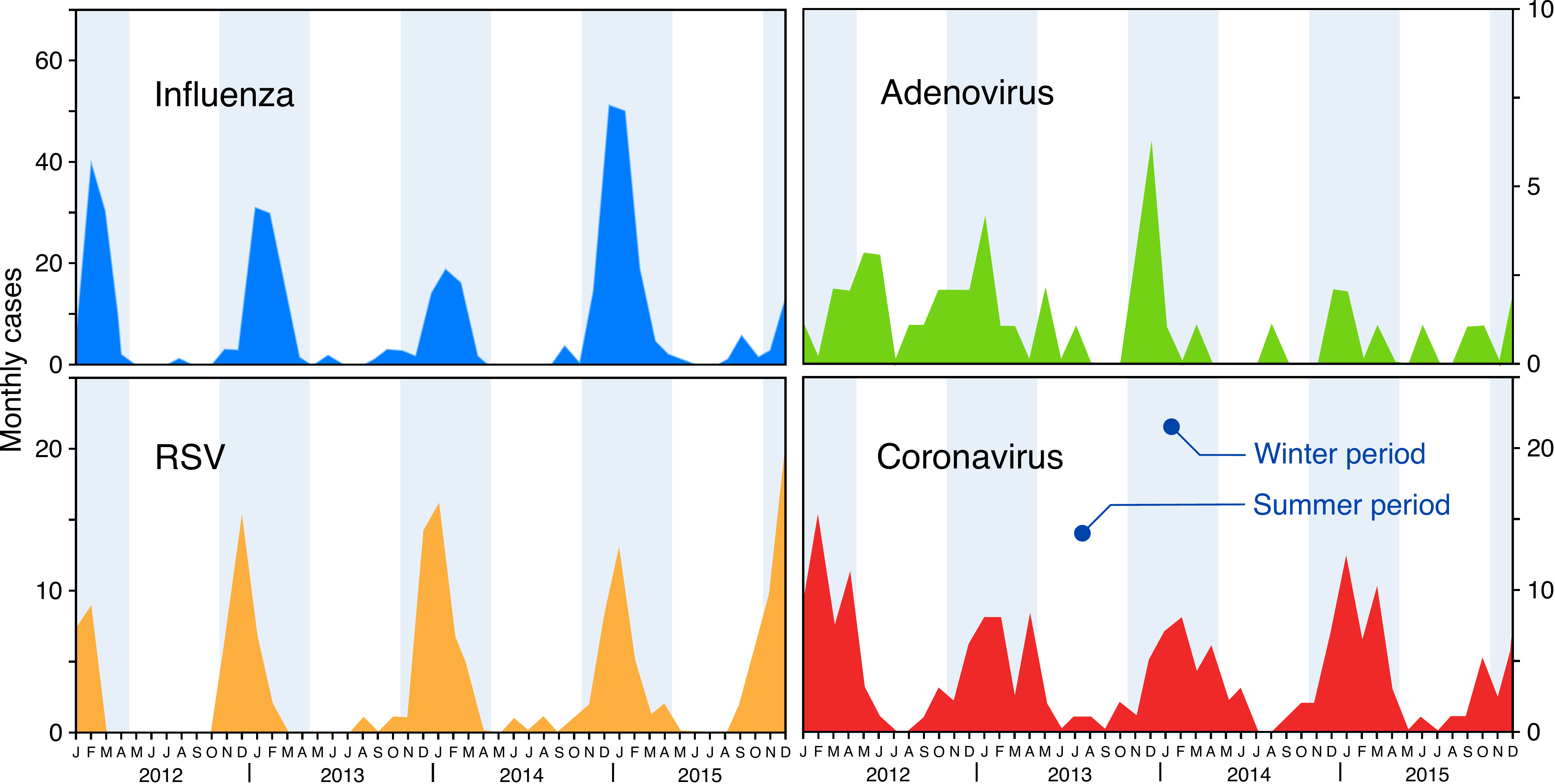}
\caption{Seasonality of four different respiratory viruses (influenza virus, adenovirus, respiratory syncytial virus (RSV), and coronavirus) in Paris during years 2012 to 2015. Shaded regions show the winter period (November to April). Image adapted from Ref.~\cite{Visseaux2017}.}
\label{fig:1}
\end{figure}

Droplet spread is the main mode of transmission for respiratory viruses such as influenza, common-cold viruses, and some SARS-associated coronaviruses, including SARS-CoV-2~\cite{Booth2005,Chan2011,Dancer2020,Verreault2008,Sooryanarain2015,Morawska2020}. A typical and very common feature of respiratory infections is {\em seasonality}, a periodic upsurge in infection incidence corresponding to seasons or other calendar periods. In fact, in temperate regions, most---but not all---human respiratory pathogens exhibit an annual increase in incidence each winter, with variations in the timing of onset and magnitude of the increase~\cite{Dowell2004,Galanti2019,Moriyama2020,Visseaux2017}, as shown in \Fig~\ref{fig:1} for several respiratory viruses. For instance, ``flu season'' in cold winter months is such a widespread and familiar phenomenon that we typically do not wonder why influenza viruses appear to have a greater reproduction rate when it is cold outside, even though they circulate year-round~\cite{Fisman2012seasonality}. On the other hand, tropical countries have much weaker annual climate cycles, and outbreaks show less seasonality and are more difficult to be explained by environmental correlations~\cite{Viboud2006,Deyle2016,Pica2012,Sooryanarain2015,Tamerius2011,Tang2014}. Yet even though the recognition of seasonal patterns in infectious diseases dates back to the era of Hippocrates, the underlying mechanisms are still not well understood~\cite{Fisman2012seasonality,Yang2011}. 

\begin{figure}[!b]
\centering
\includegraphics[width=0.5\linewidth]{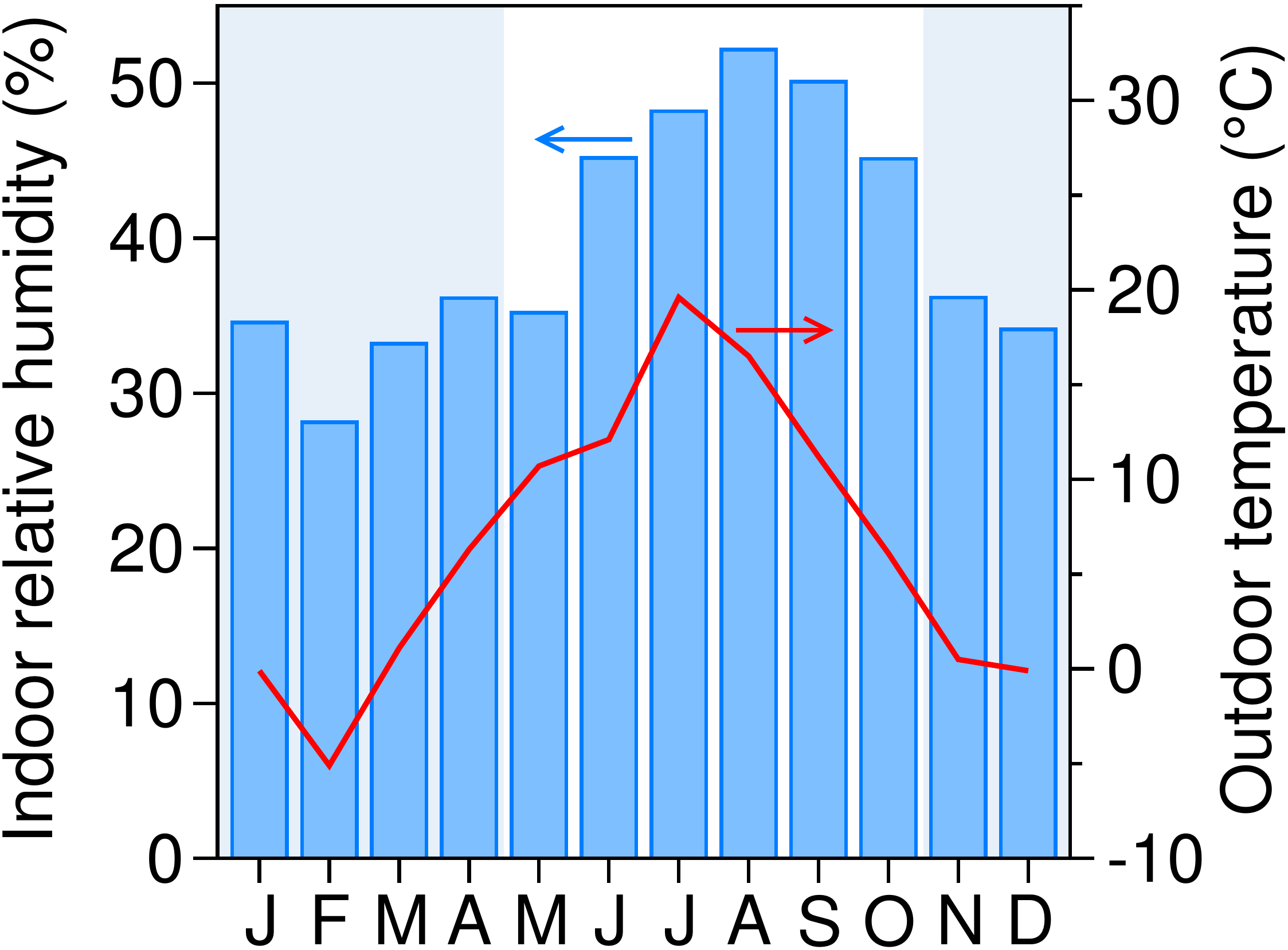}
\caption{Indoor RH (blue bars; left scale) and outdoor temperature (red line; right scale) over the period of one year in a residential building. Shaded region shows the winter period (November to April). Data for Gothenburg, Sweden~\cite{Engvall2005,temperature}.
}
\label{fig:2}
\end{figure}

Numerous factors have been suggested to drive the distinct seasonality of various pathogens: Human behavior (staying indoors more in colder months, \rev{less frequent ventilation}, school schedules)~\cite{Tamerius2011,Earn2012effects}, human immune function (diminished daylight and its impact on vitamin D metabolism~\cite{Dowell2001seasonal}), ultraviolet radiation~\cite{Harper1961}, and indoor relative humidity (RH)---the ratio of the partial pressure of water vapor to the equilibrium vapor pressure of water at a given temperature. \rev{The role of RH will be discussed in detail in this review.} One may of course wonder what the role of temperature is, as this is an environmental parameter that clearly correlates with seasons in temperate regions. However, there is little scientific evidence to suggest that lower winter temperatures are important direct drivers of wintertime seasonality of respiratory infections~\cite{Fisman2012seasonality}. In particular, in indoor environments, where people spend $90\%$ of their time and where most infections occur~\cite{Moriyama2020,Nazaroff2016}, temperature does not vary much since buildings are heated as it gets cooler outdoors~\cite{Yang2011}. Nevertheless, what the outdoor temperature does indirectly influence is the RH inside buildings. Heating the buildings in winter dries the cold air coming in from the outside, causing RH to drop dramatically. As a result, indoor RH in temperate regions typically varies between $10\%$ and $40\%$ in the winter months, which is significantly lower compared with its range of $40\%$ to $60\%$ in the summer months~\cite{Engvall2005,Hodgson2004,Nguyen2014,Zhang2010}. By way of example, \Fig~\ref{fig:2} shows the mean RH variation (blue bars) over the period of one year in a Swedish residential building: As winter turns to summer, mean indoor RH increases from $30\%$ to $50\%$, and clearly correlates with outdoor temperature (red line). In tropical regions, on the other hand, indoor RH is significantly higher throughout the year~\cite{Moriyama2020,Tariku2015,Zhang2010}.

Research shows that RH plays a paramount role in the spread of infections through a number of different mechanisms. First, RH directly impacts how and to what extent the exhaled human droplets can spread through the air. Second, the stability of winter viruses trapped in those droplets shows a striking correlation with low RH ($20\%$ to $50$\%), while the stability of summer or year-round viruses is enhanced at higher RH ($80$\%)~\cite{Moriyama2020}. And finally, dry air dries out the mucous membrane in the nose, which eases the invasion of infectious viruses into the respiratory tract~\cite{Sunwoo2006physiological,Hildenbrand2011rhinitis,Makinen2009}. A better understanding of the role of RH for virus viability in droplets thus not only helps us understand the droplet spread of infections and seasonality of some viruses, but can also guide our understanding of using humidity as a non-pharmaceutical intervention~\cite{Reiman2018,Hobday2013}.

This review aims to summarize those physical mechanisms of droplet and airborne transmission of disease in which RH plays a role. In particular, we address the question of why and how the difference in RH between $30\%$ and $50\%$ (\Fig~\ref{fig:2}) influences the spread of respiratory disease. By first estimating the droplet size and composition (\Sec~\ref{sec:size}), we use simple approximations to determine the general physics of a falling droplet (\Sec~\ref{sec:physics}) and look at how it is influenced by RH and by both the presence of solutes as well as efflorescence effects. We then separately discuss how RH impacts the sedimentation of larger droplets (\Sec~\ref{sec:sedimentation}) and the deposition of aerosol (\Sec~\ref{sec:aerosol}). Finally, we look into virus-laden droplets (\Sec~\ref{sec:viruses}) and the various factors that influence the survival of viruses in these droplets with respect to changes in RH.

\section{Respiratory droplet size and composition}
\label{sec:size}

\begin{figure*}[!b]
\centering
\includegraphics[width=0.6\textwidth]{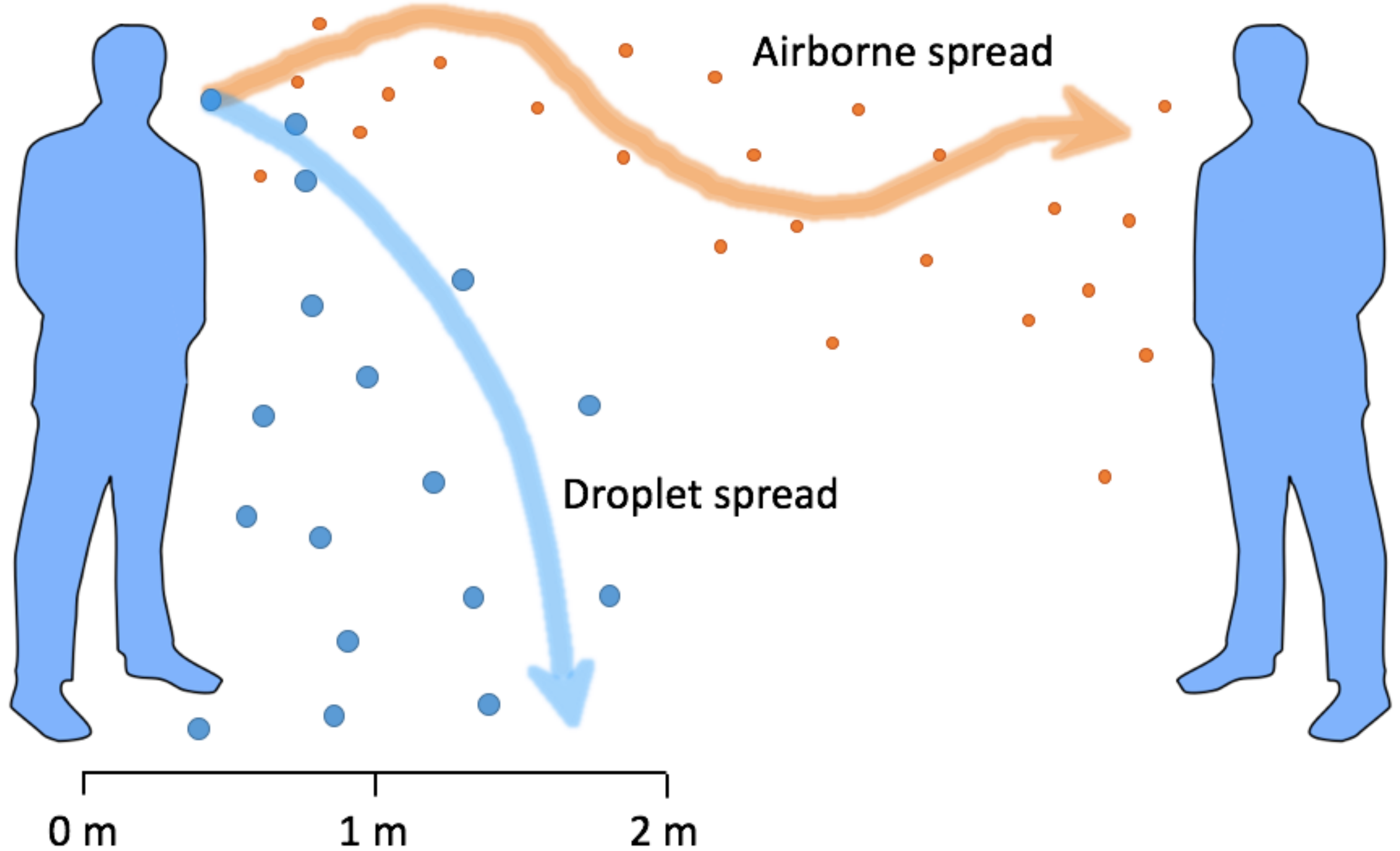}
\caption{Droplet and airborne spread of respiratory droplets. The size of the droplets influences whether they sediment quickly (droplet/fomite spread) or persist in the air for a longer period of time (airborne spread).}
\label{fig:3}
\end{figure*}

The size of droplets expelled during various human activities such as breathing, talking, singing, coughing, and sneezing is an important factor in determining their fate---whether they evaporate, sediment, or persist in the air~\cite{Belser2010,Gralton2011,Nazaroff2016,Tellier2009,Thomas2013}. Often, a distinction is made between larger respiratory {\em droplets}, which do not spread far from their origin and quickly sediment onto neighboring surfaces, potentially contaminating them and thus facilitating transmission of droplet-borne disease, and smaller {\em aerosol} particles, which are small enough to persist in the air, are influenced by various kinds of airflow, and can potentially transmit disease over larger distances (\Fig~\ref{fig:3}). While most definitions of droplets and aerosol distinguish them by a certain size cutoff---current WHO guidelines put this at $5$~$\mu$m~\cite{world2007infection}---these have varied over time, reflecting the fact that the transition between droplet and aerosol behavior is continuous rather than sharp~\cite{Gralton2011}. Most often, the term {\em droplet transmission} is defined as the transmission of disease by respiratory droplets that tend to settle quickly to the ground (typically within $1$ to $2$ m from the site of generation). Conversely, {\em airborne transmission} is defined as disease transmission by particles that are much smaller in size and can remain suspended in air for prolonged periods of time \rev{(\ie, aerosol)} and consequently travel much greater distances~\cite{Gralton2011,morawska2005droplet,xie2007far}. Furthermore, the size of respiratory droplets also influences where in the respiratory tract they can deposit~\cite{Tellier2009,Thomas2013}, and by that the severity and spread of a disease. For the purposes of this review, we will refer to all particles produced by respiratory activity as {\em droplets}, regardless of their size. We will examine both the regime when they are large enough to quickly sediment as well as the regime when they are small and are transmitted as aerosol particles.

\begin{figure*}[!t]
\centering
\includegraphics[width=0.4\textwidth]{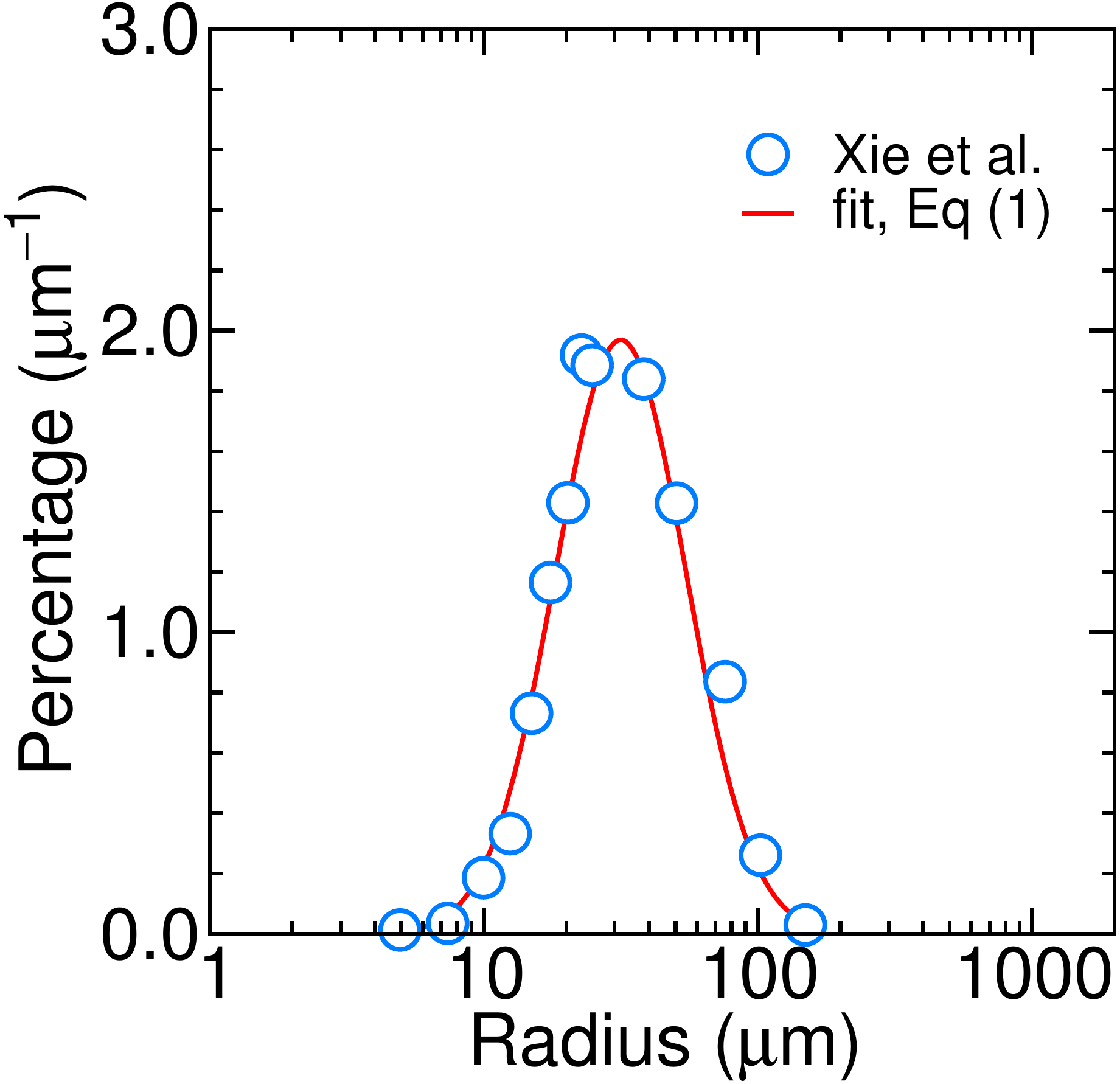}
\caption{Droplet size distribution while talking. Blue circles show measurements by Xie \ea~\cite{xie2009exhaled}, performed at $\rh=70\%$, and the solid red line shows the fit of \Eq~\eqref{eq:P}.
}
\label{fig:4}
\end{figure*}

Different human respiratory activities in general produce different amounts of droplets of varying sizes~\cite{Asadi2019,Morawska2009,Thomas2013}. \Figure~\ref{fig:4} shows an example of a size distribution of deposited droplets produced by talking, obtained in experiments of Xie \ea~\cite{xie2009exhaled}. The distribution can be fitted well by a Gaussian curve in the lin-log scale:
\begin{equation}
P(R)\sim \exp\left[-c\left(\ln\,R/R_*\right)^2\right].
\label{eq:P}
\end{equation}
The radii of the vast majority of droplets in \Fig~\ref{fig:4} are in the range of $R\sim10$ to $100$ $\mu$m, and only a minority of them have size below $10$~$\mu$m. Nonetheless, the latter importantly contribute to the airborne route of transmission, as we shall see in later sections. Size distribution of droplets does not depend greatly on the activity that produces them~\cite{Chao2009,Morawska2009,xie2009exhaled}, although droplet particles originating in the lower respiratory tract in general tend to be smaller than the particles produced in the upper respiratory tract~\cite{Johnson2011}. Some studies have also reported multimodal size distributions, which has been explained in terms of different physiological production mechanisms~\cite{Johnson2011,somsen2020small}.

While the size distributions of exhaled droplets produced by various activities are mostly similar, the activities differ greatly in the number of droplet particles produced, spanning several orders of magnitude~\cite{Fernstrom2013,Fiegel2006}. In general, more sensitive methods developed recently show that all human respiratory activities produce more droplets than previously thought~\cite{Anfinrud2020,Stadnytskyi2020}. Talking, for instance, produces over $\sim10^2$ particles per second~\cite{Anfinrud2020,Asadi2019,xie2009exhaled}, while a single sneeze can produce upwards of $10^4$ particles~\cite{Fernstrom2013}. Thus, several minutes of talking can still produce as many droplets as a single cough or a sneeze~\cite{Asadi2019,Fernstrom2013}. The number of exhaled droplets also varies between individuals, with some emitting an order of magnitude more than the others---so-called superspreaders~\cite{Asadi2019,Lloyd2005,Shen2004}. \rev{Different activities also influence the composition of the exhaled droplets (based on the location of secretion), which further varies both among as well as within individuals and is also influenced by their health, resulting in a wide range of measured concentrations of proteins, salt ions, and lipids in respiratory droplets (Table~\ref{tab:composition}).}

\begin{table}[!htp]
\caption{Basic composition of respiratory droplets of various origins.}
\label{tab:composition}
\begin{center}
\begin{tabular}{lll}
\hline\noalign{\smallskip}
\multicolumn{1}{c}{component} & \multicolumn{1}{c}{concentration range} & \multicolumn{1}{c}{references} \\
\noalign{\smallskip}\hline\noalign{\smallskip}
proteins & $\lesssim0.01$ to $80$ mg/ml & \cite{Effros2002,Scheideler1993,Spicer1984,Gould2001,Ruocco1998,reynolds1984respiratory,Sanchez2011,Vejerano2018,Marr2019} \\
salt ions & $50$ to $150$ mM & \cite{Effros2002,Vejerano2018,Marr2019} \\
lipids & $0.01$ to $30$ mg/ml & \cite{Spicer1984,Larson1980,Vejerano2018,Marr2019} \\
\noalign{\smallskip}\hline
\end{tabular}
\end{center}
\end{table}

\section{Physics of a falling droplet}
\label{sec:physics}

After droplets are expelled from the mouth or nose into the air, they undergo various physical and chemical processes---evaporation being the most notable among them---that change their structural properties in an important way. In the air, these droplets (or droplet particles) are subject to gravity, Brownian motion, electrical forces, thermal gradients, electromagnetic radiation, turbulent diffusion, and so on~\cite{morawska2005droplet}. This enormous repertoire of phenomena that accompany droplets in the air leads to a variety of possible outcomes, and our ambition is not to summarize them all in a single article. In this section, we will \rev{focus instead} on the essential physics that is necessary to explain the basics of a falling droplet, its sedimentation, and airborne spread, as well as the ways in which RH impacts this behavior.

\subsection{Evaporation of a falling water droplet}

\begin{figure*}[!t]
\begin{center}
\begin{minipage}[b]{0.60\textwidth}\begin{center}
\includegraphics[width=\textwidth]{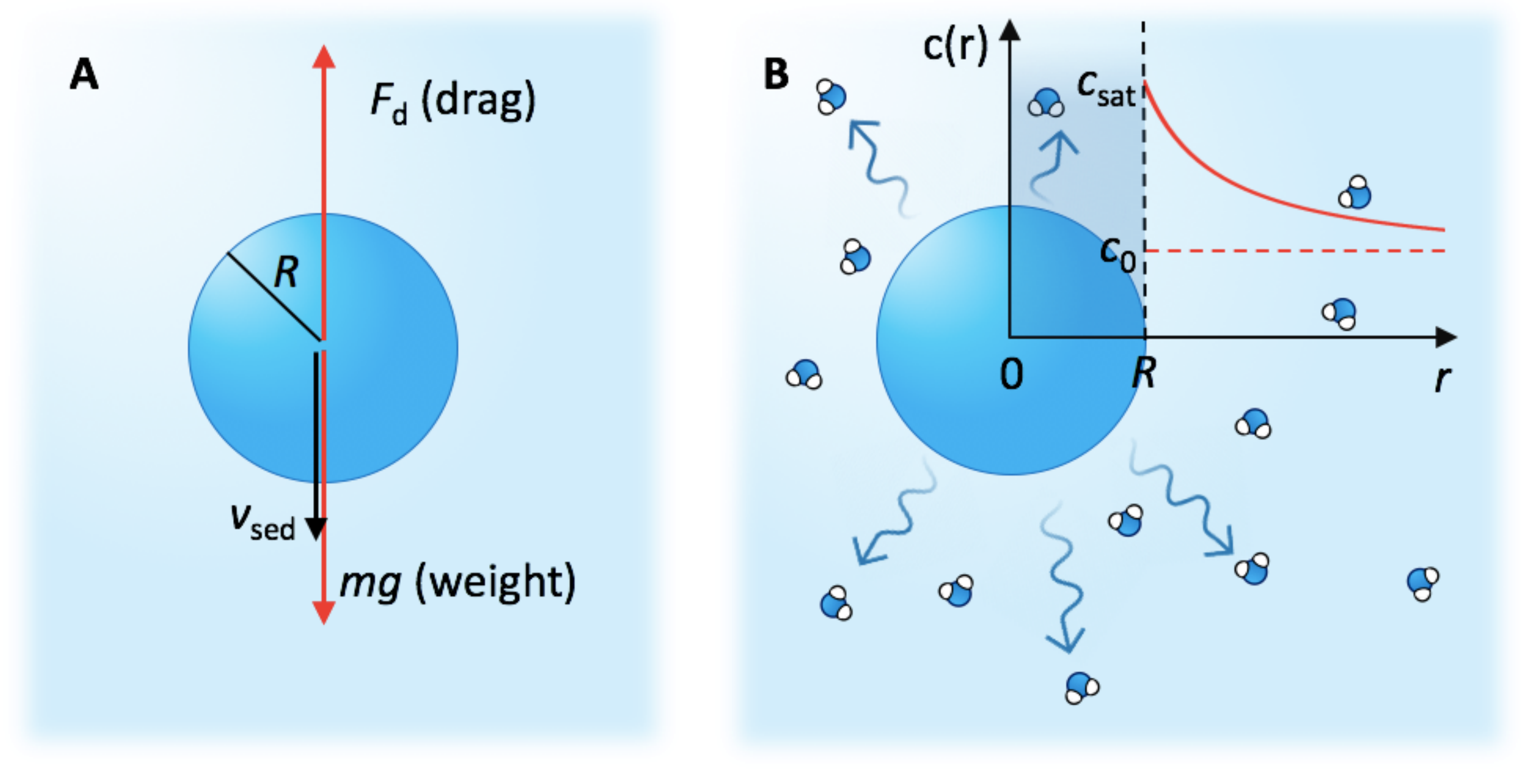}
\end{center}\end{minipage}\hspace{1ex}
\begin{minipage}[b]{0.38\textwidth}\begin{center}
\includegraphics[width=\textwidth]{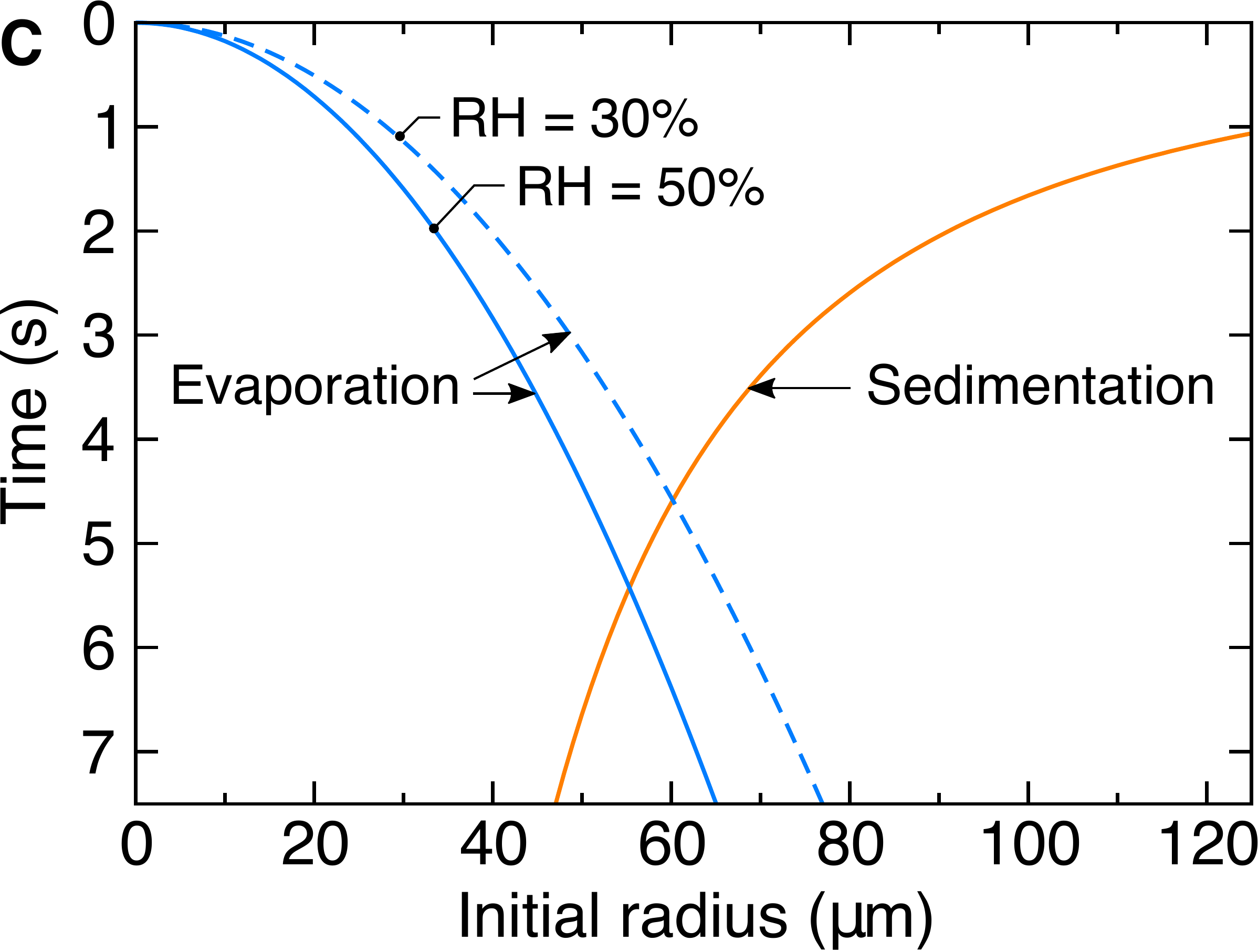}
\end{center}\end{minipage}
\caption{{\bf (A)} Droplet falling through the air with a constant, sedimentation (terminal) velocity $v_\trm{sed}$. The gravitational force is balanced by the Stokes drag. {\bf (B)} Evaporating droplet stagnant in air. The vapor density profile (inset) is given by \Eq~\eqref{eq:cr}. {\bf (C)} Wells falling-evaporation diagram, showing evaporation [blue lines; \Eq~(\ref{eq:tsed})] and sedimentation [orange line; \Eq~(\ref{eq:tev})] times of droplets as a function of the initial radius of the droplet.
}
\label{fig:5}
\end{center}
\end{figure*}

We first consider a pure water droplet of radius $R_0$ in air (\Fig~\ref{fig:5}A). Because of the gravitational force $F_\trm{g}=(4\pi/3) \rho_\trm{w}gR_0^3$ acting on the droplet (where $\rho_\trm{w}$ is the water density and $g$ the gravitational acceleration), it starts to accelerate downwards. This motion is opposed by air drag, given by the Stokes law $F_\trm{d}=6\pi\eta R_0 v$ (applicable in the regime of low Reynolds numbers), where $v$ is the droplet velocity and $\eta$ the air viscosity. Soon, the drag balances out the gravity ($F_\trm{d}=F_\trm{g}$) and the droplet reaches sedimentation (terminal) velocity of
\begin{equation}
v_\trm{sed}=\xi R_0^2
\label{eq:v}
\end{equation}
where $\xi=2g\rho_\trm{w}/9\eta\approx 1.2\times 10^{8}~\trm{m}^{-1}\trm{s}^{-1}$. Sedimentation velocities for different droplet sizes are given in \Tab~\ref{tab:1}, where we can see that they span six orders of magnitude as droplet size decreases from $100$ to $0.1$~$\mu$m. Acceleration time needed for a droplet to reach terminal velocity is $t_\trm{acc}=v_\trm{sed}/g$, which amounts to $t_\trm{acc}\approx0.1$~s for a $100$-$\mu$m-large droplet with $v_\trm{sed}=1$~m/s. Since respiratory droplets are mostly smaller than $100$~$\mu$m, this means that they quickly reach their terminal velocities and we can thus neglect any acceleration effects.

\begin{table}[!b]
\begin{center}
\setlength{\tabcolsep}{1ex}
\caption{Initial droplet radius $R_0$, sedimentation (terminal) velocity $v_\trm{sed}$ [\Eq~\eqref{eq:v}], sedimentation time $t_\trm{sed}$ [\Eq~\eqref{eq:tsed}] from the height of $h=2$~m in the absence of evaporation and in an undisturbed atmosphere \rev{(\ie, without air currents)}, and evaporation time $t_\trm{ev}$ [\Eq~\eqref{eq:tev}] at 50\% RH.
}
\begin{tabular}{rrrr}
\hline\noalign{\smallskip}
\multicolumn{1}{c}{$R_0$}	&  \multicolumn{1}{c}{$v_\trm{sed}$} & \multicolumn{1}{c}{$t_\trm{sed}$ (no evaporation)} & \multicolumn{1}{c}{$t_\trm{ev}$ ($\rh=50\%$)} \\
\noalign{\smallskip}\hline\noalign{\smallskip}
$100$ $\mu$m & $1$ m/s & $2$ s	& $20$ s\\
$10$ $\mu$m & $1$ cm/s & $3$ min & $0.2$ s\\
$1$ $\mu$m & $1$ mm/s & $5$ h & $2$ ms\\
$0.1$ $\mu$m & $1$ $\mu$m/s & $23$ days & $20$ $\mu$s \\
\noalign{\smallskip}\hline
\label{tab:1}
\end{tabular}
\end{center}
\end{table}

If a droplet is released from height $h$, the time it takes for it to reach the ground (sedimentation time) in an undisturbed atmosphere and without evaporation (\ie, it falls as a rigid body) is simply $t_\trm{sed}=h/v_\trm{sed}$. Using \Eq~\eqref{eq:v}, this reads
\begin{equation}
t_\trm{sed}=\frac{h}{\xi R_0^2}\quad\textrm{(no evaporation)}.
\label{eq:tsed}
\end{equation}
The sedimentation time is inversely proportional to the square of the droplet size and again spans orders of magnitude for the most typical respiratory droplet sizes (\Tab~\ref{tab:1}). Note that for droplets smaller than a few microns, airflow considerably disturbs the actual deposition to the ground, as we will discuss in \Sec~\ref{sec:aerosol}.

As soon as the droplet enters unsaturated air, it starts to evaporate and its radius shrinks with time. We will demonstrate by a simple calculation that evaporation plays an essential role in falling droplets. To that end, we assume a motionless droplet with respect to the surrounding air, which defines the stagnant-flow approximation~\cite{netz2020droplets}. Water molecules that evaporate from the surface of the droplet undergo diffusion in the surrounding air. Thus, the vapor number density $c$ around the droplet can be described by the diffusion equation $\partial c/\partial t=D\nabla^2 c$. The steady-state solution ($\partial c/\partial t=0$) in spherical geometry yields 
\begin{equation}
c(r)=c_0 + (c_\trm{sat}-c_0)\frac{R}{r},
\label{eq:cr}
\end{equation}
which fulfils two relevant boundary conditions: The concentration approaches the ambient vapor concentration $c(r\to\infty)=c_0$ far away from the droplet on the one hand and the saturation value at the droplet surface $c(R)=c_\trm{sat}$ on the other. The latter condition is valid in the so-called {\em diffusion-limited} regime in which evaporation is limited by the speed at which water molecules diffuse away from the droplet, creating ``free space'' for new water molecules to evaporate. Only for droplet radii below $70$~nm does the evaporation process cross over into the {\em reaction-limited} regime in which the limiting factor is the rate at which water evaporates from the droplet surface~\cite{netz2020droplets}. Typical respiratory droplets thus fall well into the diffusion-limited regime, described by \Eq~\eqref{eq:cr}.

Evaporation flux density can be obtained from Fick's first law of diffusion as $j=-D\,\rmd c(r)/ \rmd r\,|_{r=R}$. The total flux is then $J=4\pi R^2j=4\pi RD\, c_\trm{sat}\,(1-\trm{RH})$, where we have used the definition of relative humidity $\trm{RH}=c_0/c_\trm{sat}$. This now allows us to use the relation $J=-c_\trm{w}\rmd V/\rmd t$ to connect the droplet volume $V=(4\pi/3) R^3$ with evaporation time, where $c_\trm{w}=33$~nm$^{-3}$ is the number density of liquid water. From here, we can derive the time-dependent radius of an evaporating droplet
\begin{equation}
R(t)=R_0\sqrt{1-\frac{t}{t_\trm{ev}}}\quad\textrm{(pure water droplet)},
\label{eq:Rt}
\end{equation}
where $R_0$ is the initial radius. The droplet vanishes completely at the evaporation time
\begin{equation}
t_\trm{ev}=\frac{R_0^2}{\theta(1-\trm{RH})},
\label{eq:tev}
\end{equation}
where $\theta=2D\,c_\trm{sat}/c_\trm{w}=1.1\times 10^{-9}~\trm{ m}^2\trm{s}^{-1}$. The estimates for the evaporation times of droplets with different initial sizes are given in \Tab~\ref{tab:1}: As we can see, a $100$-$\mu$m-large droplet takes several seconds to evaporate, whereas a $10$-$\mu$m-large droplet evaporates in just a fraction of a second.

\Equation~\eqref{eq:tev} is, of course, only an approximation. Evaporation is a process accompanied and influenced by various phenomena~\cite{barrett1988growth,beard1971wind,davis1982transport,frossling1938uber,kinzer1951evaporation}, and detailed overview of those can be, for instance, found in work by Xie \ea~\cite{xie2007far} and Netz~\cite{netz2020droplets,NetzEaton2020}. Some of the most notable effects are {\em (i)}~evaporation cooling, where owing to the large evaporation enthalpy of water, droplet surface cools down by $\approx10$ K at $\rh=50\%$~\cite{netz2020droplets}, which in turn decreases the evaporation rate and the diffusion coefficient~\cite{xie2007far}; {\em (ii)}~Stefan flow, an induced flow of air away from the droplet caused by the evaporated vapor, which increases the evaporation rate~\cite{kukkonen1989interdependence}; {\em (iii)}~ventilation effects, where the airflow around the falling droplet speeds up evaporation, relevant for droplets larger than a few tens of microns~\cite{kukkonen1989interdependence,netz2020droplets}; and {\em (iv)}~presence of solutes, which lowers the chemical potential of water and the rate of evaporation~\cite{netz2020droplets,NetzEaton2020}. Also, once the droplet radius is smaller than $70$~nm, evaporation switches over to the reaction-limited regime and the radius decreases linearly with time~\cite{netz2020droplets,NetzEaton2020}. Any lipids present \rev{in the droplets~\cite{Spicer1984,Vejerano2018,Larsson1996}} complicate things further, since a lipid layer can be formed on its surface, significantly slowing down the evaporation~\cite{frenkiel1965evaporation,redrow2011modeling}. Nevertheless, for our purposes, \Eq~\eqref{eq:tev} is sufficient to estimate the droplet evaporation time and demonstrate the effect of RH.

We thus arrive at the question of what happens with a falling water droplet---does it fall to the ground or does it evaporate before reaching it? The answer to this question was first provided by William F.\ Wells in his seminal work in 1934~\cite{wells1934air}, where he established what we call today Wells evaporation-falling curves, shown in \Fig~\ref{fig:5}C. These curves are diagrams of time (traditionally in the reverse sense) versus the initial droplet size. Orange line in \Fig~\ref{fig:5}C shows the time needed for a droplet to reach the ground (\ie, the sedimentation time of \Eq~\eqref{eq:tsed} in absence of evaporation), and blue lines show the evaporation times at different RH [\Eq~\eqref{eq:tev}]. The diagram demonstrates a clear dichotomy of two distinct fates that depend on the initial size of the droplet: Small droplets will evaporate before reaching the ground, whereas larger ones will reach the ground before they disappear. With this diagram, Wells suggested that large respiratory droplets ($R>60$~$\mu$m) settle on the ground quickly whereas smaller ones ($R<60$~$\mu$m) dry out and any nonvolatile materials (including bacteria and viruses) stay suspended in the air for significant periods of time.

This notion provided the first clue about the difference between the transmission of infections by deposition of large droplets and by airborne routes. Wells evaporation-falling curves depend on ambient RH and suggest that higher RH slows down evaporation and increases the amount of droplets that deposit on the ground. But while Wells curves are historically important, they oversimplify the actual fate of respiratory droplets. One of the main reasons is that respiratory droplets never evaporate entirely but instead only to around half their initial size, as we will see next.

\subsection{Droplets containing solutes}

\begin{figure*}[!t]
\centering
\includegraphics[width=0.8\textwidth]{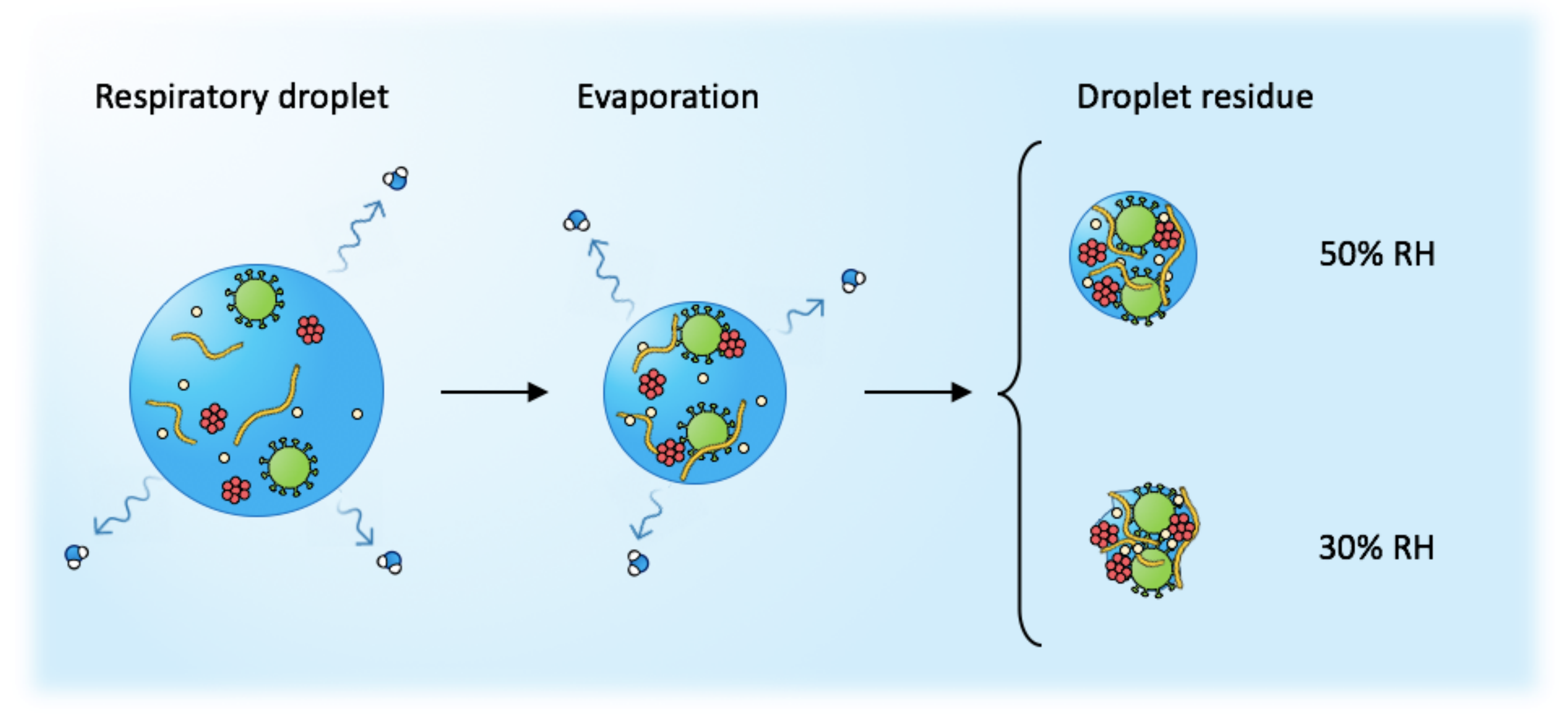}
\caption{Drying out of a respiratory droplet. The droplet progressively evaporates and shrinks in size. Its final size depends on the amount of non-water solutes in the droplet and on the ambient RH---the larger the RH, the larger the final size of the droplet residue.}
\label{fig:6}
\end{figure*}

\rev{Human respiratory droplets are composed mainly of water ($\sim90\%$ to $99\%$), with the remainder being mostly inorganic ions, sugars, proteins, lipids, DNA, and, potentially, pathogens. While the exact droplet composition depends strongly on many factors, the typical mass or volume proportion of non-water content in a droplet is around $\phi_0=1\%$ to $10\%$~\cite{Chartier2009,Effros2002,Johnson2011,Raphael1989,Spicer1984}.} A hypothetically completely dried-out droplet devoid of water would thus have a radius of 
\begin{equation}
R_\trm{dry}=R_0\,\phi_0^{1/3}\quad\textrm{(completely dried-out droplet, RH = 0)},
\label{eq:dry}
\end{equation}
between $22\%$ and $46\%$ of the initial droplet radius. However, a droplet never dries out completely as some water remains sorbed inside, its amount governed by RH~\cite{nicas2005toward}. Currently, it is unclear how droplet composition influences the final droplet size and the response to RH~\cite{liu2017evaporation}. Clearly, respiratory droplets contain very complex organic macromolecular structures, made out of mostly hydrophilic molecules with considerable hydration effects, which are consequently strongly hygroscopic~\cite{liu2017evaporation}. We will examine two extreme cases of the response of droplet size to RH (\Fig~\ref{fig:6}). In the first case, we will assume ideal mixing of solutes with water, which allows for a simple mathematical derivation. The second case involves crystallization of salt in the droplet, triggering an abrupt change in the droplet size. 

\subsubsection{Droplet size: Ideal mixing}
\label{ssec:ideal}

From a thermodynamic perspective, evaporation occurs because the liquid water in the droplet has a higher chemical potential than the unsaturated vapor phase surrounding the droplet. The chemical potential of the vapor phase at a given $\rh$ relative to bulk water (or saturated vapor at $\rh=100\%$) is $\Delta \mu_\trm{v}\approx \kB T\,\ln\,\trm{RH}$, where $\kB$ is the Boltzmann constant and $T$ the temperature. When a nonvolatile solute is introduced into the droplet, the water chemical potential subsides. In the approximation of ideal mixing, the chemical potential of water with solute is smaller than the one of pure water by $\Delta \mu_\trm{w}\approx \kB T\,\ln\,x_\trm{w}$, with $x_\trm{w}$ the mole fraction of water. If the initial amount of solute in the droplet is small ($x_\trm{w}\lesssim1$) and the surrounding relative humidity is $\rh<$100\%, water starts to evaporate from the droplet. With that, $x_\trm{w}$ decreases over time, and so does the chemical potential. Evaporation continues until the chemical potential of water in the droplet reaches the one of vapor, $\Delta\mu_\trm{v}=\Delta\mu_\trm{w}$. From here, we obtain an estimate that the evaporation stops once the water fraction in the droplet reaches $x_\trm{w}\approx\trm{RH}$ and, likewise, when the solid content is $x_\trm{s}=1-x_\trm{w}\approx1-\trm{RH}$. \rev{Note that the contribution of the Laplace pressure to the chemical potential is negligibly small and becomes relevant only for nanometer-sized droplets.}

The final, equilibrium volume of the dried out droplet, termed the {\em droplet residue} or {\em droplet nucleus}, is the sum of the water and solute content. For simplicity, we assume that solute molecules are of similar size as the water molecules, which makes the total volume of the droplet residue larger than the volume of a completely dried out droplet [\Eq~\eqref{eq:dry}] by a factor of $1+x_\trm{w}/x_\trm{s}=1/(1-\rh)$, namely~\cite{netz2020droplets}
\begin{equation}
R=R_0\left(\frac{\phi_0}{1-\trm{RH}}\right)^{1/3}\quad\textrm{(droplet residue)}.
\label{eq:Rresidue}
\end{equation}
If we \rev{again} assume the solute content of respiratory droplets to be between $\phi_0=1\%$ and $10\%$, the dried-out droplet shrinks down to $27\%$ to $58\%$ of the initial radius at $\trm{RH}=50\%$. In other words, a droplet residue is between $1/4$ and $1/2$ of the initial size of an exhaled respiratory droplet, as has already been suggested by numerous studies~\cite{nicas2005toward,Yang2011,liu2017evaporation}. RH thus controls not only the evaporation rate [\Eq~\eqref{eq:tev}] but, as implied by \Eq~\eqref{eq:Rresidue}, also the final size of the residue (as noted by, \eg, Effros \ea~\cite{Effros2002} and Nicas \ea~\cite{nicas2005toward}). An important consequence is that at higher RH, droplet residues are larger, which in turn makes them sediment to the ground faster. According to \Eq~\eqref{eq:Rresidue}, increasing RH from $30\%$ to $50\%$ increases the residue size by $(0.7/0.5)^{1/3}\approx10\%$. This result is, of course, very approximate and merely provides qualitative insights. More complex mathematical models take into account the non-ideality of mixing and even distinguish between aqueous salt phase and the insoluble solid material (\eg, mucous organics and potential pathogens)~\cite{liu2017evaporation}. The qualitative conclusions are, however, always the same: Droplet nuclei maintain a larger size in humid air than in dry air. Nonetheless, the precise relationship between the size of a respiratory droplet residue and RH is still largely unknown, and research on how RH influences the final size of droplet residues is surprisingly very limited~\cite{liu2017evaporation}.

\subsubsection{Droplet size: Non-ideal mixing and efflorescence}
\label{ssec:efflorescence}

\begin{figure*}[!t]
\begin{center}
\begin{minipage}[b]{0.33\textwidth}\begin{center}
\includegraphics[width=\textwidth]{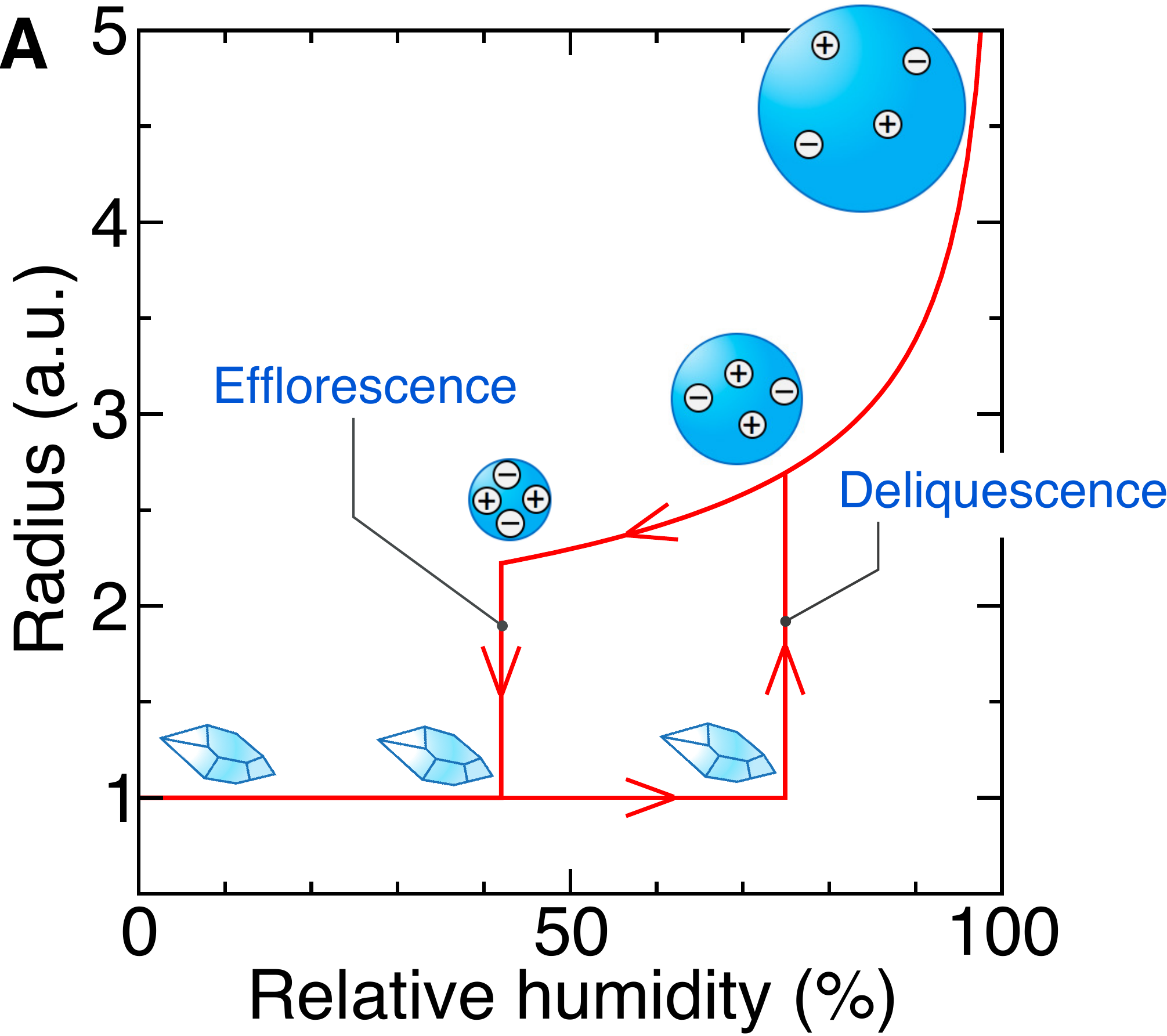} 
\end{center}\end{minipage}
\begin{minipage}[b]{0.32\textwidth}\begin{center}
\includegraphics[width=\textwidth]{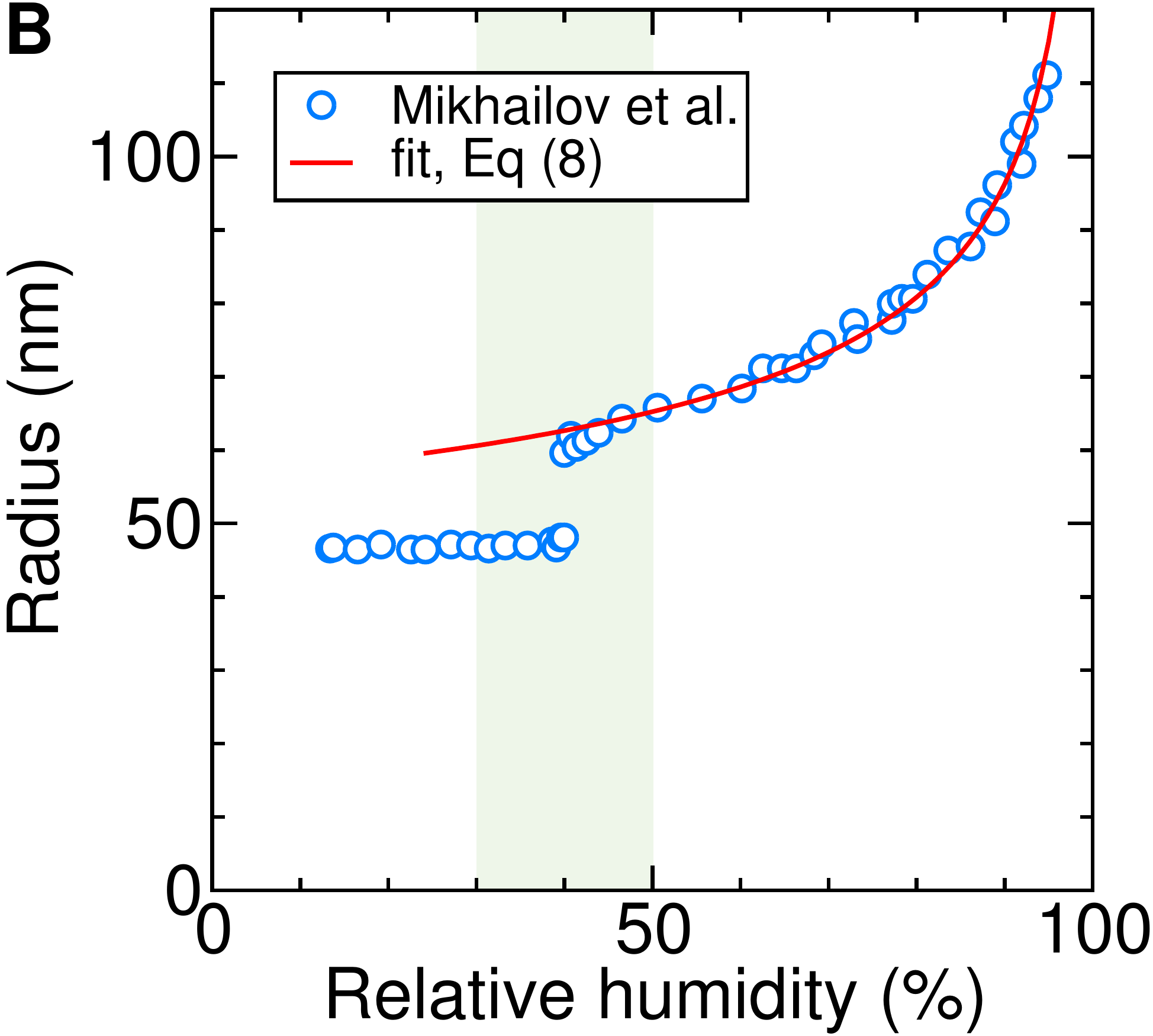} 
\end{center}\end{minipage}
\begin{minipage}[b]{0.30\textwidth}\begin{center}
\includegraphics[width=\textwidth]{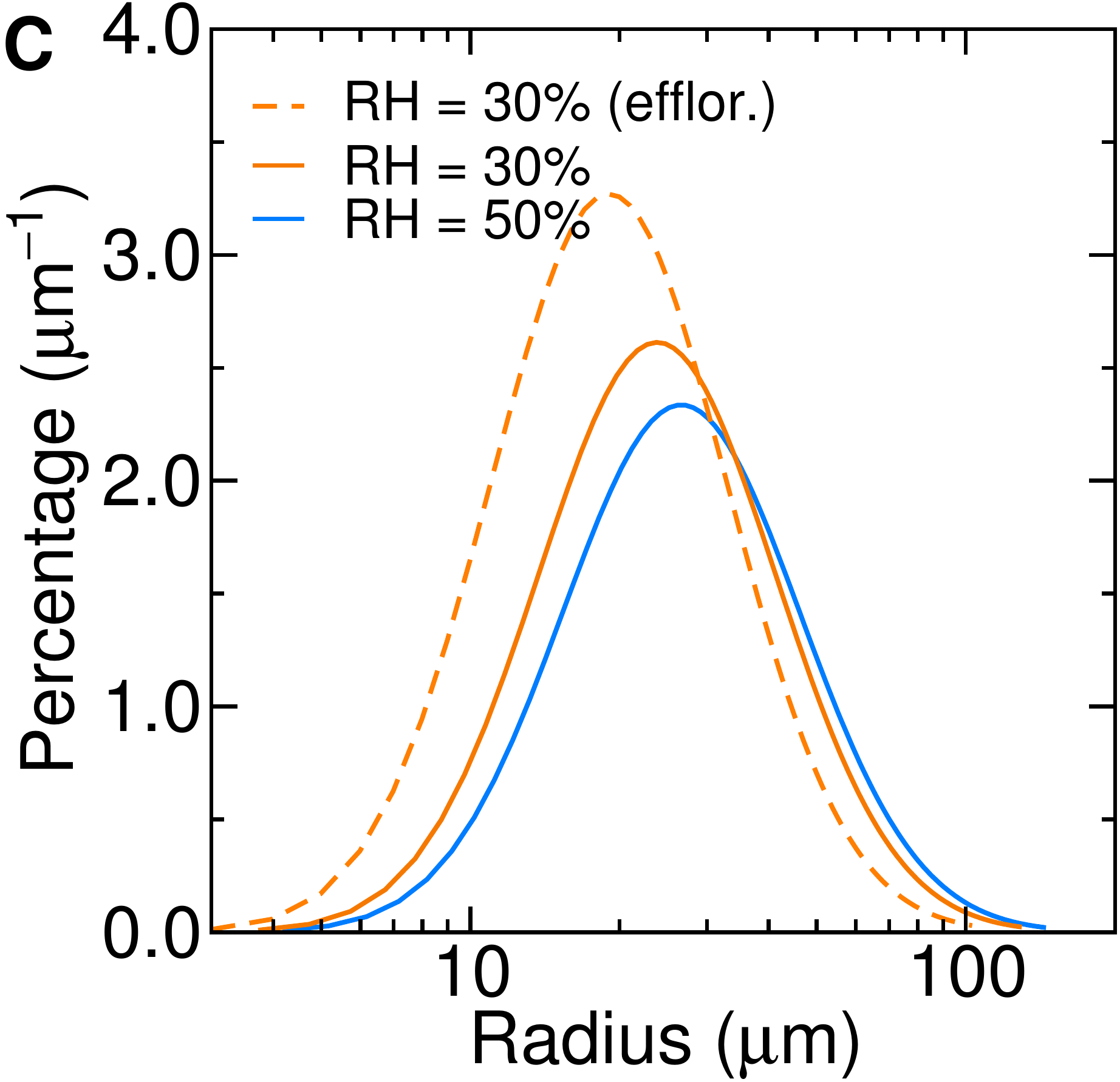} 
\end{center}\end{minipage}
\caption{
{\bf (A)} Illustration of hygroscopic shrinking and growth of a NaCl-water droplet. Upon dehydration, a liquid droplet undergoes efflorescence at $\rh\approx40\%$ and crystallizes, whereas upon hydration, a NaCl particle undergoes deliquescence at $\rh=75\%$ and turns into a liquid droplet. {\bf (B)} Radius of NaCl-water droplets containing bovine serum albumin protein with an initial dry mass fraction of $10\%$, measured upon dehydration (Mikhailov et al.~\cite{Mikhailov2004}). The red dashed line is a fit of \Eq~\eqref{eq:Rresidue} to the data points for $\rh>50\%$. The green-shaded region denotes the most typical ambient conditions of $\rh=30\%$ to $50\%$. {\bf (C)} Size distribution of droplet residues at different RH, recalculated from \Fig~\ref{fig:4} using \Eq~\eqref{eq:Rresidue}.
}
\label{fig:7}
\end{center}
\end{figure*}

Non-ideal mixing of solutes with water can lead to dramatic, non-continuous processes in a shrinking droplet during evaporation. For instance, \rev{during evaporation of water containing inorganic salts such as NaCl, salt concentration increases, and in the bulk phase, salt crystallization typically occurs upon reaching saturation. However, in small droplets, salt concentration can} overcome the solubility limit and can push the system deep into a supersaturated, metastable state~\cite{Gregson2018}. If evaporation continues (\eg, if RH is low enough) and salt concentration increases further, salt eventually crystallizes, and all the water evaporates in an {\em efflorescence} transition\rev{---solidification of supersaturated solution---}as shown in \Fig~\ref{fig:7}A. The inverse process, termed {\em deliquescence} \rev{(liquefaction of a solid particle)}, occurs at significantly higher RH.

\rev{The metastability can typically be explained by classical nucleation theory as an interplay between the interfacial free energy of the crystal-water phase boundary and the chemical potential difference between the water and crystalline phases. The interfacial free energy acts as a barrier to the formation of a crystal nucleus, which enables droplets to reach very high levels of supersaturation before efflorescence occurs.
Efflorescence in small droplets can proceed via different pathways, ranging from homogeneous nucleation, internally mixed nucleation (relevant for droplets with multiple components that can act as nucleating agents), and contact-induced nucleation (due to collisions between droplets)~\cite{pohlker2014efflorescence, davis2017crystal}. Exact details are, however, not entirely understood~\cite{davis2017crystal}. }
Whether or not efflorescence transition occurs depends both on the type of salt~\cite{Martin2000,Mikhailov2004} as well as on the size and composition of the droplet~\cite{Biskos2006,Cheng2015,Martin2000}. For instance, NaCl has an efflorescence RH of $\approx40\%$~\cite{choudhury2013pattern,Cohen1987,Gregson2018,Mikhailov2004}, while no transitions are observed for ammonium nitrate~\cite{Mikhailov2004}.

Are these crystallization phenomena also important for respiratory droplets? The majority of our knowledge on hygroscopic properties of small particles comes from atmospheric sciences~\cite{martin2008phase,o2015liquid,posfai1998wet,you2012images,marcolli2006phase}, which is unfortunately difficult to generalize to respiratory droplets, as the latter contain a different and typically much more complex organization of organic compounds such as proteins and lipids~\cite{kumar2003organic,Mikhailov2004,mokbel1997study}. \rev{Furthermore, droplet composition varies widely between individuals and within individuals over time~\cite{Effros2002}, especially in the case of disease~\cite{pacht1991deficiency,Sanchez2011,Spicer1984}, and so does the response of the droplets to RH.} A good starting point for understanding the hydration nature of respiratory droplets are thus simple salt-water droplets containing proteins. As an example, \Fig~\ref{fig:7}B shows how a droplet containing NaCl and $10\%$ of bovine serum albumin protein---a popular model for proteins and macromolecular compounds---shrinks upon dehydration when lowering the ambient RH, obtained from experiments by Mikhailov \ea~\cite{Mikhailov2004}. At high RH, droplet size follows very well the approximate expression given by \Eq~\eqref{eq:Rresidue}. At around $\rh=40\%$, however, the radius suddenly drops, indicating an efflorescence transition. Consequently, the difference in the sizes of droplet residues at $30\%$ and $50\%$ RH is almost $40\%$---much larger than predicted by the ideal-mixing model [without an efflorescence transition, \Eq~\eqref{eq:Rresidue}]. Interestingly, efflorescence of aqueous particles, whether with NaCl and/or protein content, occurs at RH of $\approx40\%$---right in the middle of the typical range of ambient RH of $30\%$ to $50\%$!  Notably, this shows that efflorescence transition can be an essential player at typical indoor conditions that govern the dynamics of respiratory droplets. Mikhailov \ea~\cite{Mikhailov2004} further observed that upon dehydration, proteins seemingly limit the nucleation of salt crystals, leading to higher stability of supersaturated salt solution and inhibition of NaCl efflorescence.

Synthetic droplets with a respiratory-fluid-like composition were also shown to undergo a phase separation upon dehydration where mucin proteins separate and localize at the droplet surface, forming an envelope~\cite{Vejerano2018}. This can inhibit access of water vapor to the particle core and lead to kinetic limitations of water exchange, phase transitions, and microstructural rearrangement processes~\cite{posfai1998wet,Mikhailov2004}. Besides surface and kinetic effects, proteins and comparable organic macromolecules can also influence the thermodynamic properties of the aqueous bulk solution, and there is even indirect evidence for changes in pH and gelation processes~\cite{Vejerano2018}. It is also important to stress that phase transitions of sub-micron particles can behave very differently than predictions made using phase diagrams of bulk materials~\cite{mifflin2009morphology}.

From this discussion, we can conclude that when RH is lowered from $50\%$ to $30\%$, a respiratory droplet residue can shrink in diameter anywhere from $\sim 10\%$ (in the case of ideal mixing) to $\sim 40\%$ (when efflorescence occurs) of its initial size. RH and efflorescence transition thus have important implications for the general distribution of respiratory droplet sizes: Using the droplet residue size distribution of Xie et al.~\cite{xie2009exhaled} at $\rh=70\%$ (\Fig~\ref{fig:4}), we can recalculate the size distributions at $\rh=50\%$ and $\rh=30\%$ using \Eq~\eqref{eq:Rresidue}, yielding the solid curves in \Fig~\ref{fig:7}C, while the dashed line shows the case of efflorescence at $\rh=30\%$ (assuming a $40\%$ shrinking compared to the sizes at $\rh=50\%$). Despite these effects, the possibility that (at least some) droplets can effloresce has typically been neglected in most---but not all---theoretical models~\cite{xie2007far,liu2017short,wang2020transport,dbouk2020coughing,Pendar2020numerical,busco2020sneezing}. In the next two sections, we will discus the implications of the droplet shrinking range (in the presence and absence of efflorescence) on the sedimentation dynamics and aerosol deposition of respiratory droplets.

\section{Droplet sedimentation}
\label{sec:sedimentation}

When assessing the risk of droplet and airborne transmission of disease, a crucial parameter is the time that respiratory droplets spend in the air before they deposit to the ground or other surfaces. We have seen in \Sec~\ref{sec:physics} that the evaporation of a typical respiratory droplet stops once it shrinks down to around half of its initial size. We can now estimate the sedimentation time of such a droplet, anticipating that RH will have a significant effect---unlike in our estimate for the sedimentation time of a droplet in the absence of evaporation [\Eq~\eqref{eq:tsed}].  We will follow the approximation recently proposed by Netz~\cite{netz2020droplets}, in which droplet radius shrinks from the initial value of $R_0$ down to the equilibrium value $R=\kappa R_0$, where $\kappa$ is a shrinking factor, given by \Eq~\eqref{eq:Rresidue} for the case of ideal mixing. The shrinking occurs in time $t_\trm{ev}^*=(1-\kappa^2)\,t_\trm{ev}$, where $t_\trm{ev}$ is the evaporation time of a pure water droplet [\Eq~\eqref{eq:tev}]. After $t_\trm{ev}^*$, droplet radius remains constant at $\kappa R_0$. With this assumption in mind and using \Eqs~\eqref{eq:v} and \eqref{eq:Rt}, we can write the expression for time-dependent sedimentation velocity as
\begin{equation}
v_\trm{sed}(t)=\xi R_0^2
\left\{
\begin{array}{ll}
1-t/t_\trm{ev}&;\quad t\leq t_\trm{ev}^*\\
\kappa^2 &;\quad t>t_\trm{ev}^*
\end{array}\right..
\label{eq:vsedt}
\end{equation}
Since we can neglect acceleration effects (\Sec~\ref{sec:physics}), we integrate the velocity in \Eq~(\ref{eq:vsedt}) up to the time when the droplet touches the ground. This gives us the height $h$ from which the droplet was released,
\begin{equation}
h=\xi R_0^2
\left\{
\begin{array}{ll}
t-t^2/\,2t_\trm{ev}&;\quad t\leq t_\trm{ev}^*\\
\kappa^2\, t+(1-\kappa)^2\, t_\trm{ev}/2&;\quad t>t_\trm{ev}^*
\end{array}\right..
\label{eq:h}
\end{equation}

\Equation~\eqref{eq:h} provides the relationship between the initial droplet radius and its sedimentation time, and allows us to make a simple estimate of how quickly respiratory droplets settle to the ground at various RH. If we know the total number of pathogen particles contained within the droplets in the air---referred to as the {\em pathogen load}---we can estimate the concentration of the pathogen remaining in the air at a given time (\eg, after a single exhalation event, such as a cough or a sneeze). Assuming further that pathogen concentration is the same in all exhaled droplets, the initial pathogen load (\ie, the total number of pathogen particles) is simply proportional to the total initial volume of the droplets exhaled into the air. Evaporation alone does not change the number of pathogens in the droplets, and we can thus evaluate the {\em relative pathogen load} $f_\trm{load}$ (the load at a given time relative to the initial load) as the cumulative volume of the (dried-out) droplet residues that have not yet sedimented to the ground divided by the volume of all initially exhaled droplet residues. To do this, we integrate the (dried-out) droplet volume weighted by a suitable size distribution $P(R)$ from $0$ up to a cutoff radius $R_\trm{sed}(t)$, the upper radius of those droplets that have not yet sedimented during time $t$. Thus, $R_\trm{sed}(t)$ is obtained as $\kappa R_0$ from \Eq~\ref{eq:h}. The relative load is then
\begin{equation}
f_\trm{load}(t)=C\int_0^{R_\trm{sed}(t)} \frac{4\pi R^3}{3}\times P(R)\rmd R.
\label{eq:load}
\end{equation}
The coefficient $C$ ensures the normalization to the initial condition $f_\trm{load}(0)=1$, \ie, that at $t=0$ all the exhaled droplets are in the air.

\begin{figure*}[!b]
\begin{center}
\begin{minipage}[b]{0.35\textwidth}\begin{center}
\includegraphics[width=\textwidth]{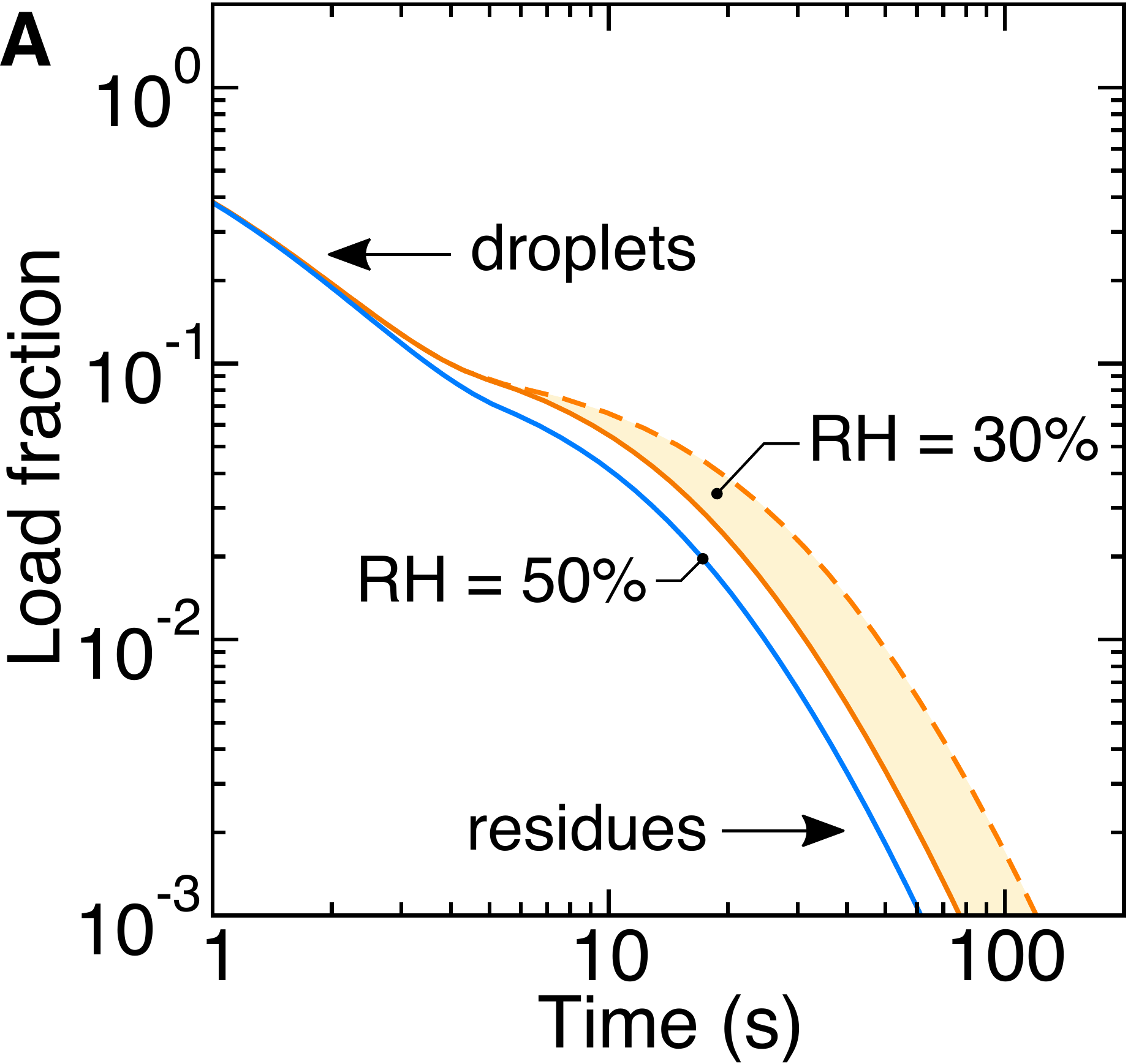}
\end{center}\end{minipage}\hspace{5ex}
\begin{minipage}[b]{0.35\textwidth}\begin{center}
\includegraphics[width=\textwidth]{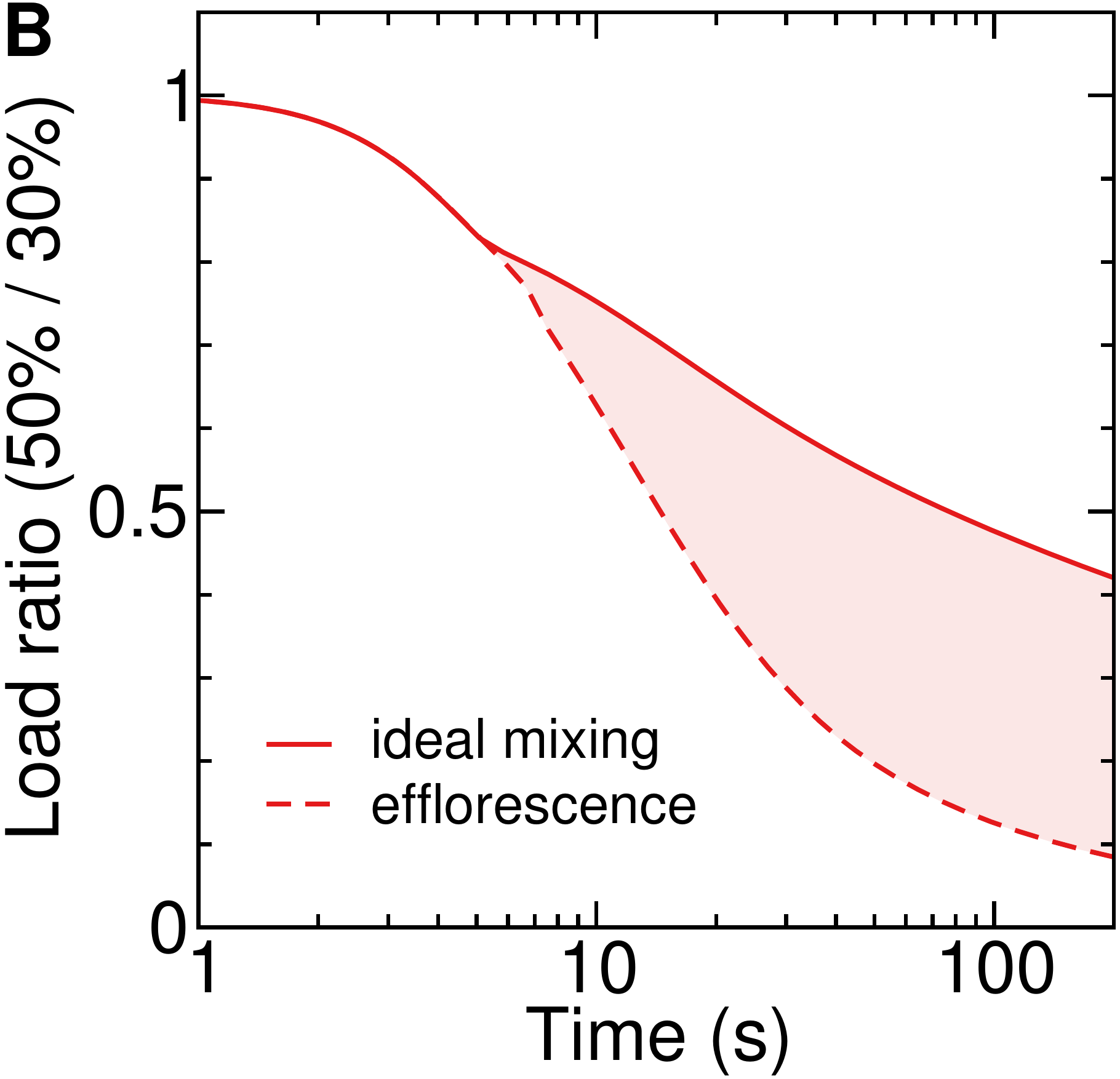}
\end{center}\end{minipage}
\caption{
{\bf (A)} Fraction of the initial pathogen load from a cough according to \Eq~\eqref{eq:load}. The distribution of  droplet residues is taken from the distributions in \Fig~\ref{fig:7}C, assuming $\phi_0=10\%$. Shown for $\rh=50\%$ and $\rh=30\%$; in the latter case both in the presence (dashed line) and in the absence (solid line) of efflorescence. {\bf (B)}~Ratio of the pathogen loads at $\rh=50\%$ and $\rh=30\%$ from panel A.
}
\label{fig:8}
\end{center}
\end{figure*}

As an example, we calculate the relative pathogen load dynamics for the three droplet size distributions $P(R)$ shown in \Fig~\ref{fig:7}C. We use \Eq~\eqref{eq:Rresidue} to calculate the initial distribution of droplet radii $R_0$ and assume that the initial volume fraction of dry material is $\phi_0=10\%$. The resulting relative loads as a function of time [\Eq~\eqref{eq:load}] are shown in \Fig~\ref{fig:8}A. We can see that the vast majority (around $90\%$) of the initial load released in the exhaled droplets from the height of $2$~m settles to the ground within $3$ to $4$~s, independently of RH. During this initial period, only the largest droplets, which did not significantly evaporate during the descent, have sedimented---they fall almost as rigid bodies~\cite{busco2020sneezing}. Afterwards, the effects of evaporation and RH become noticeable. At $\rh=50\%$ (blue curve), droplets settle more rapidly than at $\rh=30\%$ (orange-shaded band), owing to the fact that {\em (i)} evaporation is slower and {\em (ii)} the final droplet residues are larger at higher RH. At $\rh=30\%$ and in the absence of efflorescence (where the droplet residues are $10\%$ smaller than at $\rh=50\%$; shown by solid orange line), sedimentation is only marginally slower. However, sedimentation slows down significantly when efflorescence occurs (and the droplet residues are $40\%$ smaller than at $\rh=50\%$; indicated by the dashed orange line).

To express the result in another way, we can also calculate the ratio between the relative pathogen loads at $\rh=50\%$ and $\rh=30\%$, as shown in \Fig~\ref{fig:8}B. The difference gradually increases over time: After $10$~s, the relative pathogen load at $50\%$ RH is between $60\%$ to $75\%$ of that at $30\%$ RH.  After one minute, this difference already amounts to $2$- to $6$-fold. This simple calculation demonstrates that despite quite small differences in the residue sizes in the RH range of $30\%$ to $50\%$, changes in RH and the presence or absence of efflorescence still have a significant impact on the settling of droplets from the air. 

The calculation presented in this section demonstrates in a simple way how a decrease in RH from $50\%$ to $30\%$ affects the deposition of respiratory droplets. More complex models take many other hydrodynamic factors into account, such as the buoyancy effect and the speed of the exhaled air jet, ventilation, and breathing mode~\cite{xie2007far,liu2017short,wang2020transport,dbouk2020coughing,Pendar2020numerical,busco2020sneezing}. The conclusions are always qualitatively consistent with the prediction of Wells~\cite{wells1934air} that droplets settle more slowly at low RH. The studies, however, usually do not quantify the differences between $30\%$ and $50\%$ RH, relevant for ambient conditions. Moreover, most of them, as already noted, unfortunately do not take into account the possibility of salt efflorescence.

\section{Deposition of aerosol}
\label{sec:aerosol}

Our estimate of droplet sedimentation, presented in \Fig~\ref{fig:8}, implies that at $\rh=50\%$ around $99\%$ of the initially exhaled droplet volume from a typical cough settles within $\sim30$~s. The remaining droplet residues that are still in air at that point are smaller than $25$ $\mu$m in radius [see \Eq~\eqref{eq:tsed}]. Even though these small particles present less than $1\%$ of the initially exhaled pathogen load, they can travel substantially longer distances and are also more likely to be captured in the respiratory tract~\cite{mccluskey1996detection}. Moreover, infectious droplets smaller than $10$~$\mu$m can penetrate deeper into the respiratory tract (\ie, pulmonary region, which is the most sensitive part of the lungs) and have more serious health implications~\cite{Gralton2011,nicas2005toward}. Unlike larger droplet residues, settling of these small residues is not governed by gravitational sedimentation alone but is influenced also by other environmental factors~\cite{laRosa2013,morawska2005droplet,Nazaroff2016,Vuorinen2020}. Namely, the air in a typical indoor setting is never undisturbed but is overwhelmed with air currents and flows, which impact the sedimentation of the small droplets.

\begin{figure*}[!t]
\begin{center}
\begin{minipage}[b]{0.68\textwidth}\begin{center}
\includegraphics[width=\textwidth]{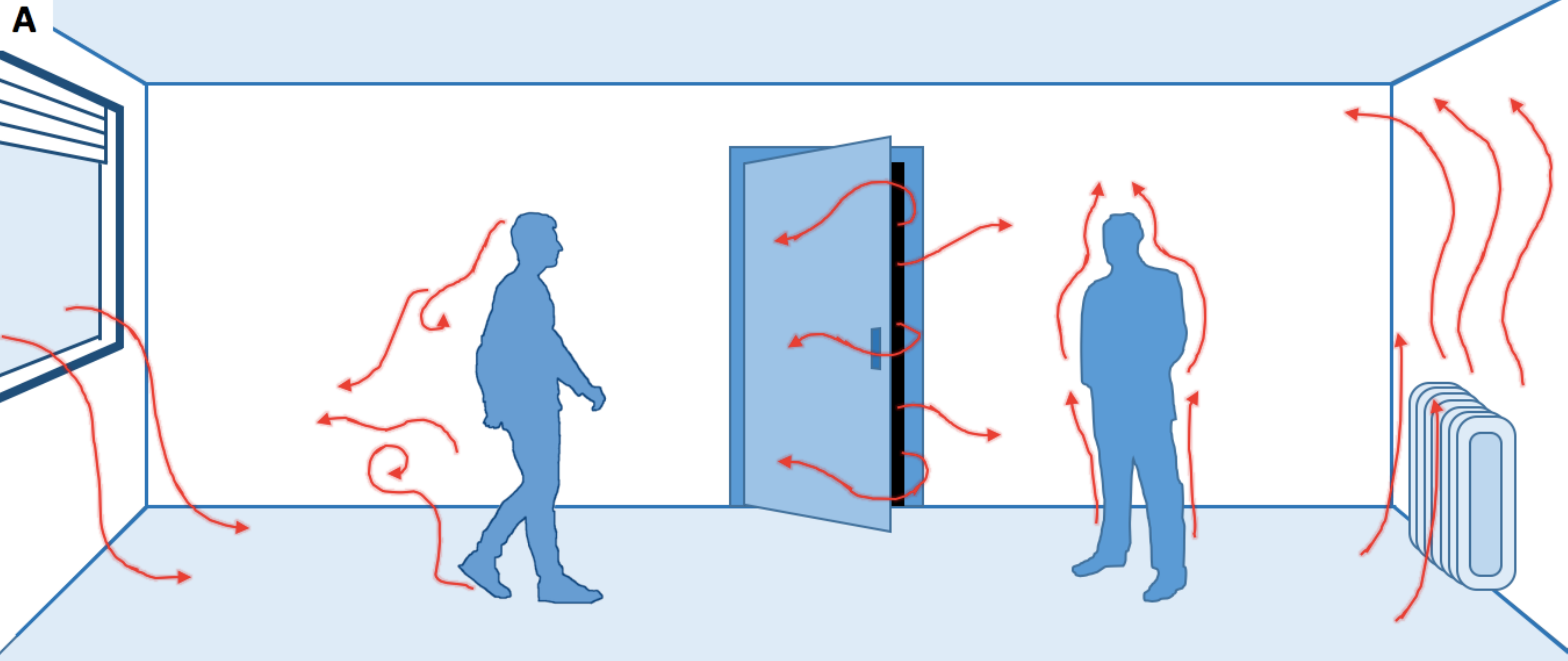}
\end{center}\end{minipage}\hspace{3ex}
\begin{minipage}[b]{0.24\textwidth}\begin{center}
\includegraphics[width=\textwidth]{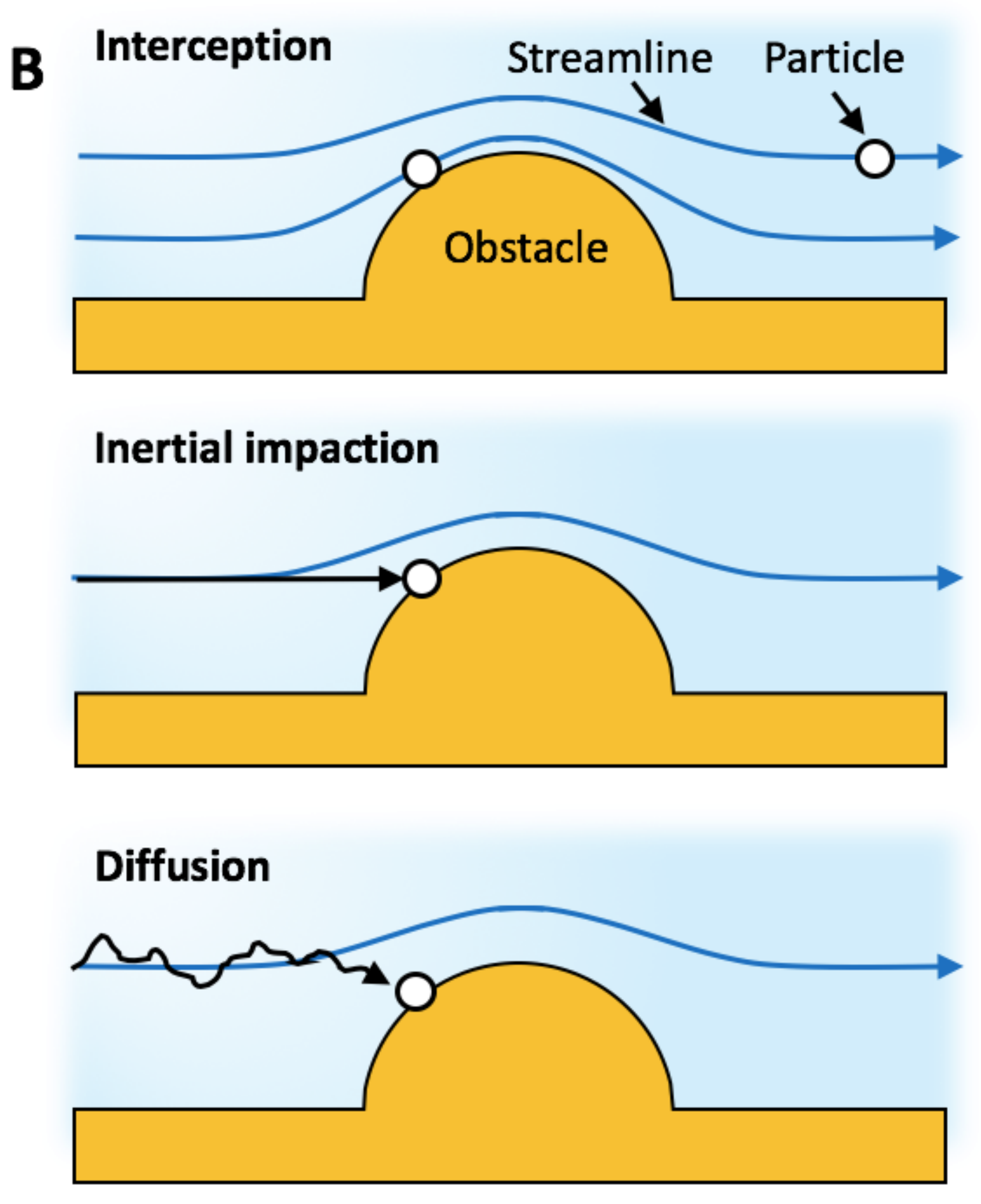}
\end{center}\end{minipage}
\caption{{\bf (A)} Airflow that influences small aerosol droplets in a typical indoor setting (from left to right): Air exchange, walking, door vortices, thermal plume, heating convection. {\bf (B)} Different deposition mechanisms of aerosol particles: Interception, inertial impaction, and diffusion. 
}
\label{fig:9}
\end{center}
\end{figure*}

Typical airflow speeds inside buildings are around $0.1$ m$/$s (convection currents, breathing, human thermal plume), but certain human activities (walking, opening and closing doors) produce short-lived flows of around $1$~m$/$s (\Fig~\ref{fig:9}A)~\cite{wei2016airborne}. These currents perturb the sedimentation of those droplets whose sedimentation velocity is comparable to or smaller than the airflow speed, which is relevant for particles smaller than $\approx20$~$\mu$m (see \Tab~\ref{tab:1}). The smaller the particle, the more it is influenced by the airflow rather than the gravity and the longer it remains in the air~\cite{morawska2005droplet}. These small respiratory droplet residues form a suspension in air and are therefore often referred to as {\em aerosol} or even {\em bioaerosol}~\cite{morawska2005droplet,Nazaroff2016}. As described in \Sec~\ref{sec:size}, the exact boundary for what should be considered aerosol is challenging to define, and the transition between the two regimes is not at all sharp, but continuous instead~\cite{Gralton2011}.

Small aerosol particles eventually collide and stick to the ground as a result of impact, which happens when the air stream changes its direction and the particle comes in contact with a surface (interception) or when the particle does not follow the changing air stream because of inertia (inertial impaction; relevant for larger particles). For sub-micron particles, Brownian diffusion is an important transport mechanism, enabling them to deviate from air streams, thus increasing their probability of deposition (\Fig~\ref{fig:9}B)~\cite{jonsson2014bioaerosol}. The crossover between interception, impaction, and diffusion deposition mechanisms lies in the range of $0.1$ to $1$~$\mu$m, where the contribution of the three mechanisms is the smallest, and aerosol particles in this size range therfore stay in the air the longest~\cite{hussein2009deposition,Nazaroff2016}.

Deposition mechanisms of aerosol are, contrary to the sedimentation of large droplets, not a deterministic but a stochastic process. In a closed indoor environment, without active filtering and in a well-mixed condition, the concentration $c_R(t)$ of particles of radius $R$ decays exponentially with time~\cite{thatcher2002effects}:
\begin{equation}
c_R(t)=c_R(0)\rme^{-(\lambda+\beta_R)\, t}.
\label{eq:depo}
\end{equation}
Here, $\lambda$ is the air-exchange rate in the room \rev{(corresponding to ventilation)} and $\beta_R$ is the first-order deposition loss rate coefficient. Air-exchange rate vanishes in a perfectly sealed room, but typically varies between $\lambda = 5$~h$^{-1}$ in residential buildings and $20$~h$^{-1}$ in public spaces~\cite{toolbox} and does not depend on aerosol type. On the other hand, the deposition rate $\beta_R$ varies widely across different conditions~\cite{thatcher2002effects}: It depends both on the aerosol properties (\eg, size and shape) as well as on the environmental parameters such as surface area, surface roughness, setup (\eg, furnishing), airflow conditions, electrical charge, and temperature gradients~\cite{byrne1995stable,long2001using,thatcher2002effects,vette2001characterization,xu1994deposition,hussein2009deposition,wang2020assessment}.

\begin{figure*}[!t]
\begin{center}
\begin{minipage}[b]{0.35\textwidth}\begin{center}
\includegraphics[width=\textwidth]{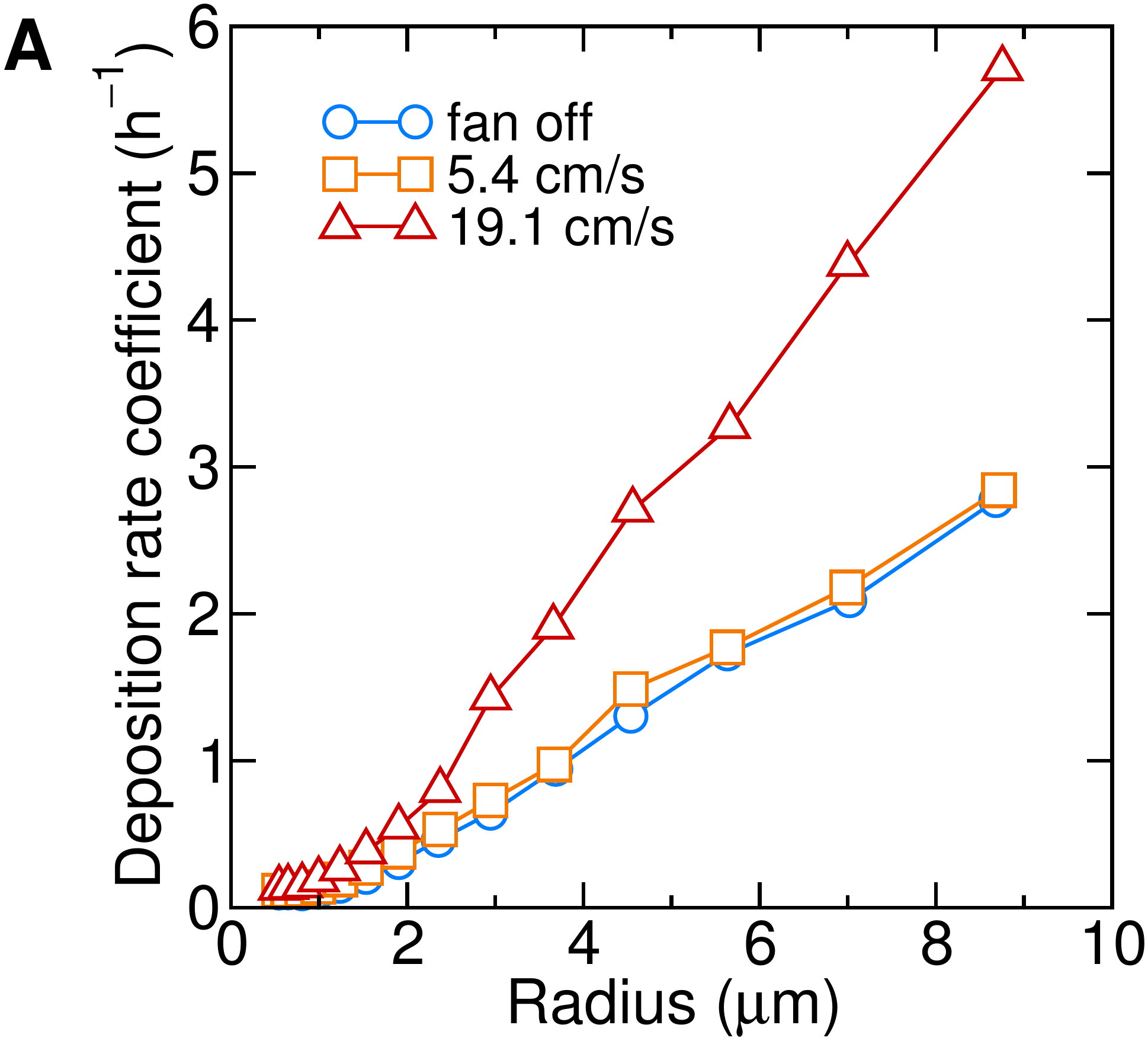}
\end{center}\end{minipage}\hspace{5ex}
\begin{minipage}[b]{0.33\textwidth}\begin{center}
\includegraphics[width=\textwidth]{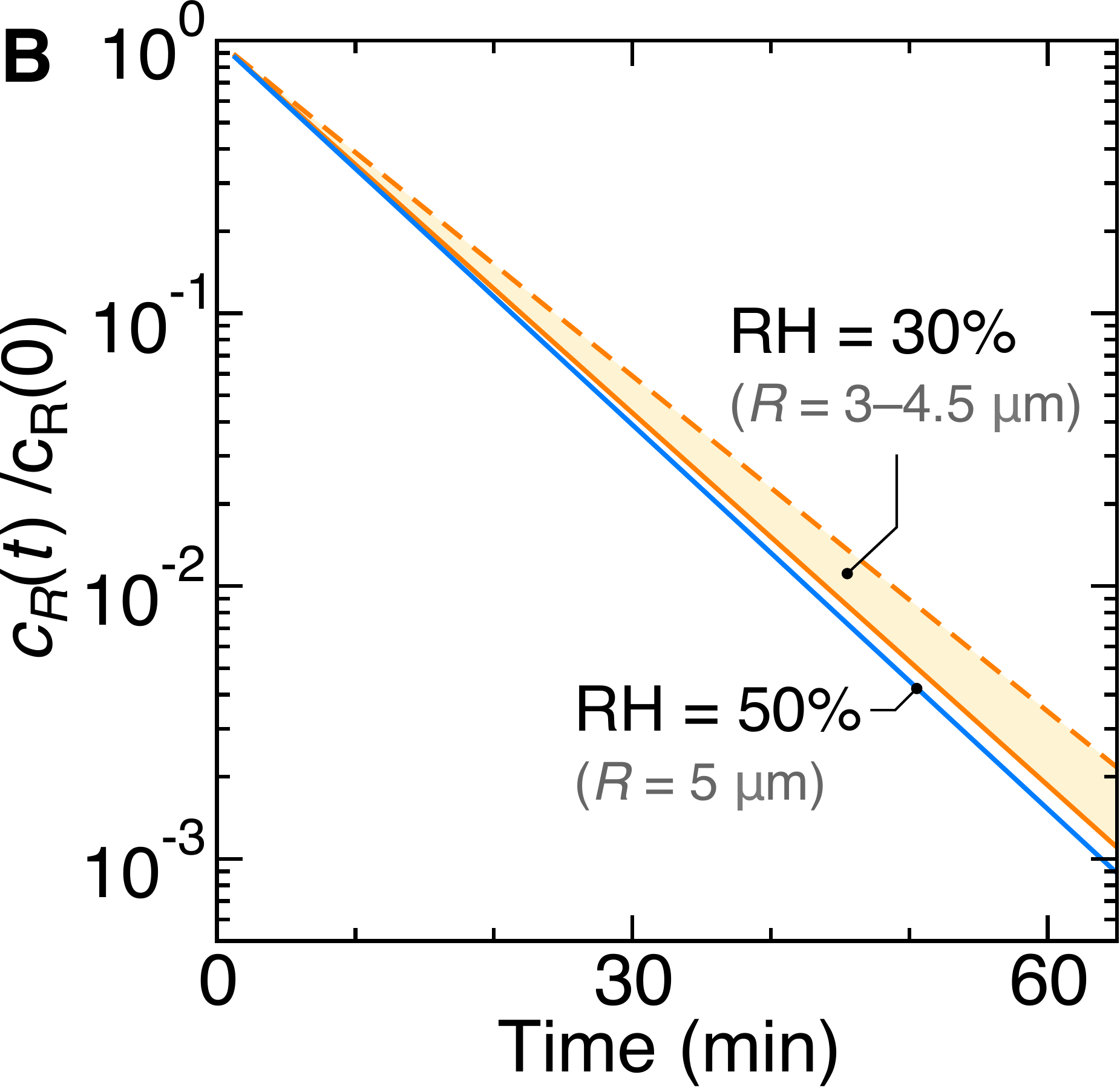}
\end{center}\end{minipage}
\caption{{\bf (A)} Particle deposition loss rate coefficient $\beta_R$ as a function of particle radius $R$, measured in a $14$ m$^3$ furnished room, either without ventilation or ventilated by fans of two different intensities (data from Ref.~\cite{thatcher2002effects}). {\bf (B)} Decay of aerosol concentration of different sizes, mimicking the effect of RH on a respiratory droplet residue with $R=5$~$\mu$m at $\rh=50\%$ (blue solid line) and $R=3$ to $4.5$~$\mu$m at $\rh=30\%$ (orange-shaded band). \rev{Calculated using \Eq~\eqref{eq:depo} with the air-exchange rate of $\lambda = 5$~h$^{-1}$}.
}
\label{fig:10}
\end{center}
\end{figure*}

\Figure~\ref{fig:10}A shows the loss rate coefficients for small particles with radii from $1$ to $10$~$\mu$m in a furnished room with several ventilation intensities (no fan, mild fan, and strong fan). We can see that the deposition rate scales roughly linearly with the particle size, $\beta\propto R$ (in this regime, interception and inertial impaction dominate over the diffusion deposition). Enhanced air movement increases the rate of particle deposition because it delivers particles more rapidly to the surfaces where they deposit~\cite{Nazaroff2016}. The mean \rev{lifetime} of aerosol in the air scales as $(\lambda+\beta_R)^{-1}$ according to \Eq~\eqref{eq:depo}. For an aerosol residue with $R=5$~$\mu$m at $\rh=50\%$ (with $\lambda=5$~h$^{-1}$ and $\beta_R=1.5$~h$^{-1}$), this makes the mean \rev{lifetime} approximately $9.2$~min. When RH is lowered to $30\%$, this particular residue shrinks down to $R=3$ to $4.5$~$\mu$m, depending on whether efflorescence transition occurs or not (\Fig~\ref{fig:7}C). Consequently, the deposition rate is reduced ($\beta_R=0.66$ to $1.3$~h$^{-1}$), and the mean \rev{lifetime} extends to $9.5$ to $11$ min. Even though this might not seem much, the difference builds up with time owing to the exponential nature of deposition, as shown in \Fig~\ref{fig:10}B. The difference is negligible in the first minutes, but after an hour, the ratio between concentrations at 50\% and 30\% RH ranges from $0.44$ to $0.82$ (the span again reflecting the presence and absence of efflorescence).

Clearly, the effect of RH on deposition of small aerosol particles is much weaker than its effect on the sedimentation of larger droplets (\Sec~\ref{sec:sedimentation}). Even though it is not entirely clear how RH influences the size of droplet residues, the physics of deposition in the two regimes is completely different. Sedimentation time of larger droplets ($R\gtrsim20~\mu$m) scales inversely with the square of their size, $t_\trm{sed}\propto 1/R^2$ [\Eq~\eqref{eq:tsed}], while the \rev{lifetime} of smaller aerosol particles ($R\sim1$ to $10$~$\mu$m) \rev{in the air} scales roughly as $\sim 1/(\lambda+\alpha R)$ [\Eq~\eqref{eq:depo}]. The size dependence of the \rev{lifetime} of smaller aerosol particles is thus much weaker than for larger droplets. Finally, for aerosol sizes in the range of $0.1$ to $1$ $\mu$m, the deposition mechanisms are very weak and measurements consequently exhibit substantial variability. Therefore, it is much more difficult to predict the fate of sub-micron aerosol particles~\cite{Nazaroff2016,thatcher2002effects}.

\section{Viruses in respiratory droplets}
\label{sec:viruses}

One of the important reasons to try to understand the behavior of respiratory droplets, their aerosolization, and sedimentation is that they often carry viruses and other pathogens and are thus an important source of disease transmission~\cite{Fernstrom2013,Kutter2018,laRosa2013,Verreault2008}. The amount of viable, infectious virus particles in an individual droplet is characterized by viral load---the amount of virus in a given volume of the droplet medium (\eg, sputum or saliva)~\cite{Gralton2011}. Recent review by Poon \ea~\cite{Poon2020} points out that even at high viral loads of $10^6$ to $10^9$ copies/ml (characteristic of, \eg, SARS-CoV-2~\cite{Bar2020,Pan2020,To2020,To2020b,Wolfel2020} and some other respiratory viruses~\cite{Wang2004,Yezli2011}), this results in approximately only $1\%$ of small, $5$-$\mu$m-large droplets carrying one or more virus particles. This is in accord with recent studies of respiratory viruses, which have shown that even large loads of viral RNA translate into only a small count of viruses in droplet and aerosol particles~\cite{Leung2020,Stadnytskyi2020} (note also that viral load given by RNA count can overestimate the number of infectious viral particles~\cite{Bar2020}). However, the sheer amount of droplets produced during various activities such as speaking or coughing (\Sec~\ref{sec:size}) can carry a significant number of viruses even when a single droplet contains at most few copies~\cite{Xu2020}. Estimates for instance show that $1$ minute of loud speaking can generate more than $10^3$ virus-containing droplets that can remain airborne for more than $8$ minutes~\cite{Stadnytskyi2020}.

As droplets are exhaled, viruses start a voyage that is all but ``hospitable''---dehydrated droplet particles can be in fact a very hostile environment. Viruses are now directly exposed to various harmful factors such as temperature~\cite{Sattar1987,Sooryanarain2015,Tang2009,Verreault2008}, UV radiation~\cite{First2007,Tseng2005,Walker2007}, atmospheric gasses~\cite{Mik1977,Tseng2006}, different types of surfaces on which droplets deposit~\cite{Chan2011,vanDoremalen2020,Kim2012,Mahl1975,Thomas2008}, and other factors~\cite{Benbough1971b,Bovallius1987,Haddrell2017,Happ1966}. It is thus not surprising that in general viruses cannot survive in these conditions indefinitely. Nonetheless, many viruses can remain infectious for long periods in both airborne and deposited droplet particles~\cite{Benbough1971b,Verreault2008}. The most significant factors influencing their survival appear to be temperature, humidity, and the nature and composition of the droplets themselves~\cite{Moriyama2020,Sattar1987,Sooryanarain2015,Tang2009,Verreault2008}. Of these, increase in environmental temperature (\eg, from $10$ to $30$~$^\circ$C) quite universally speeds up the decay of viruses in droplets~\cite{Elazhary1979a,Harper1961,Hermann2007,Ijaz1987a,Schoenbaum1990}. Note again, however, that despite this fact, low winter temperature is not considered as a direct driver of the seasonality of respiratory infections~\cite{Fisman2012seasonality}, mostly because the temperature is regulated in indoor spaces, where most infections occur~\cite{Moriyama2020,Nazaroff2016}.

\begin{figure}[!b]
\centering
\includegraphics[width=0.38\linewidth]{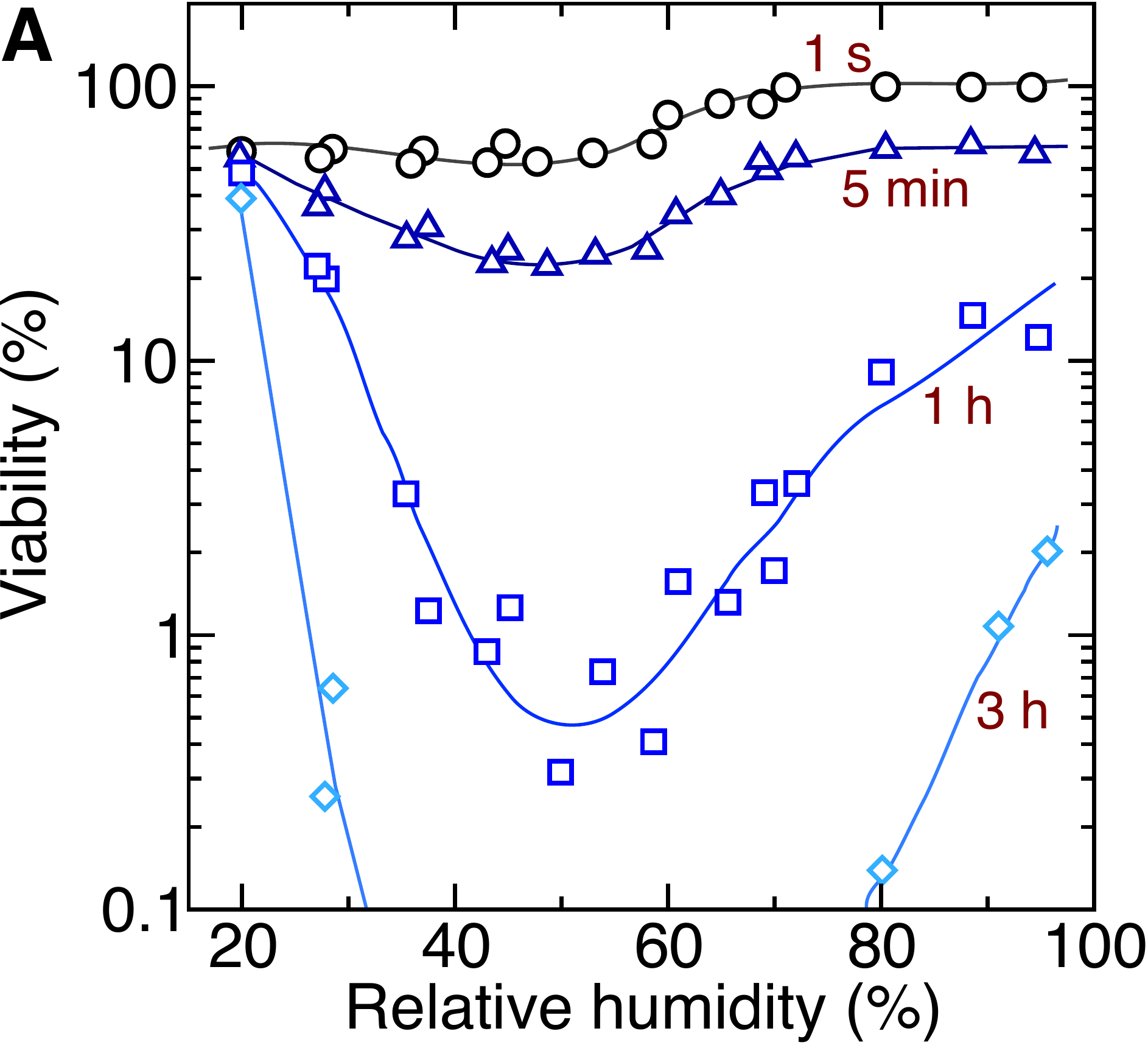}\hspace{2ex}
\includegraphics[width=0.54\linewidth]{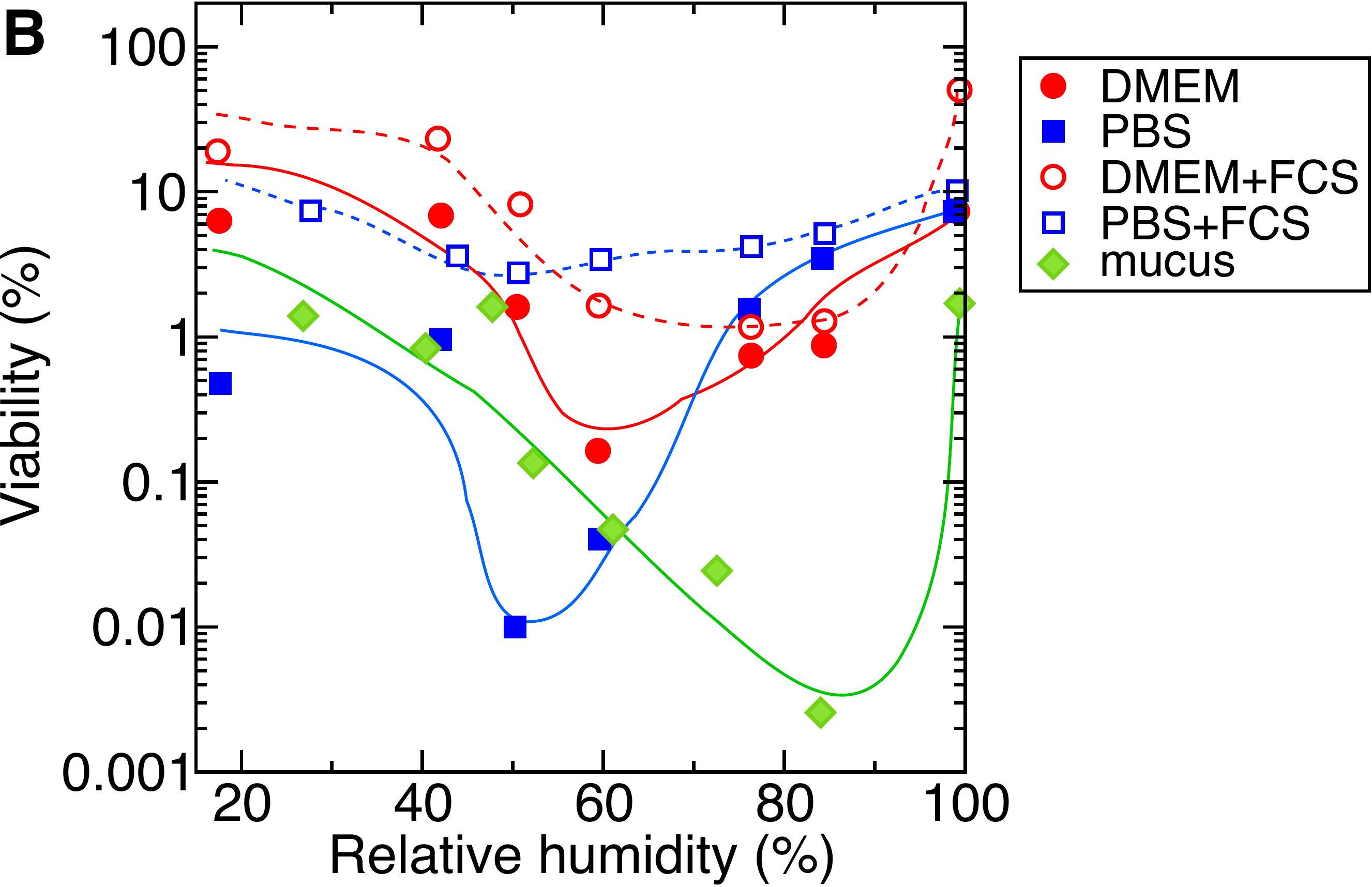}
\caption{{\bf (A)} Viability of Langat virus with respect to RH at different times after aerosolization~\cite{Benbough1971b}. {\bf (B)} Viability of influenza A virus after aerosolization in different media: Mainly salts [phosphate-buffered saline (PBS), Dulbecco's modified Eagle's medium (DMEM)], salts with addition of proteins [fetal calf serum (FCS)], and mucus~\cite{Yang2012b}. Lines are guides to the eye. }
\label{fig:11}
\end{figure}

RH, on the other hand, is more directly related to seasonal changes because of indoor heating (\Fig~\ref{fig:2}), but its impact on viruses is not universal. In particular, virus viability decreases as RH falls below $100\%$, since the droplet gets more and more dehydrated and the environment progressively deviates from physiological conditions. Surprisingly, virus viability very often recovers as RH is decreased below $50\%$, giving rise to a quite common U-shaped viability curve in response to RH, as seen in \Fig~\ref{fig:11}. In the relevant range of ambient conditions, $\rh=30\%$ to $50\%$, decreasing the air humidity can thus increase the amount of viruses that survive in the droplets. Importantly, from the infection point of view, this effect acts together with the effect of weaker droplet deposition at low RH in making drier air more effective for the spread of infections. As we will see in the following, the response of virus viability to changes in RH is affected in different ways and degrees by virus structure, the presence, composition, and concentration of solutes, pH gradients, and the available air-water interface, making it difficult to draw general conclusions~\cite{Tang2009,Verreault2008,Yang2012}.

\subsection{RH and virus structure}

Response of virus survival to changes in RH has often been linked to the presence or absence of a lipid envelope~\cite{Tang2009,Verreault2008,Yang2012}. Broadly speaking, enveloped viruses (such as influenza viruses, coronaviruses, and RSV) tend to survive longer at low RH ($\lesssim30\%$), while non-enveloped viruses (such as adenoviruses, rhinoviruses, and polioviruses) tend to survive longer at high RHs ($70\%$ to $90\%$), with $\mathrm{RH}\sim100\%$ being in general favorable for virus survival regardless of the lipid envelope. Enveloped viruses furthermore often show the distinct, U-shaped non-monotonic pattern in the response of their survival to changes in RH (\Fig~\ref{fig:11}), with a significant decrease in survival at intermediate RH~\cite{Tang2009,Verreault2008,Yang2012}. These general observations are, however, not always consistent, and exceptions abound among both enveloped and non-enveloped viruses~\cite{Donaldson1976,Ijaz1987a,Ijaz1985b,Rabey1969,Schoenbaum1990,Songer1967}. The effects of RH on survival differ from virus to virus~\cite{Lowen2007,Prussin2018,Verreault2008} or even between different strains of the same virus~\cite{Donaldson1972,Sooryanarain2015}, making it difficult to draw any conclusions based on the structure of the virus alone. For instance, Lin and Marr~\cite{Lin2019} recently studied inactivation kinetics of two bacteriophages, the enveloped $\Phi6$ and the non-enveloped MS2. They observed that the magnitude of decay was similar between the two, and both viruses showed a non-monotonic pattern in viability with respect to changes in RH, suggesting a common inactivation mechanism.

\subsection{RH and droplet composition}

Evaporation of water from the droplet induces various physico-chemical transformations in the droplet such as changes in the concentration of solutes (\eg, ions and proteins) as well as changes in the pH~\cite{Lin2019,Marr2019,Vejerano2018,Yang2012b}. Not only does droplet composition thus influence the sedimentation behavior of respiratory droplets (as we exhaustively discussed in \Sec~\ref{sec:physics}), but it also crucially affects the way an aerosolized virus survives or is inactivated~\cite{Benbough1971a,Vejerano2018,Zuo2014,Yang2012b}. While the micro-environment in the droplet is close to physiological conditions at very high RH ($\sim100\%$) and becomes dry when RH is low ($\lesssim30\%$; \Fig~\ref{fig:6}), it is likely that virus viability at intermediate RH is governed mostly by the droplet composition, giving rise to the non-monotonic U-shaped pattern of the virus survival response to RH (\Fig~\ref{fig:11}B). To understand the role that droplet composition plays in the response of virus viability in droplets at different RH, we must see how the main components of respiratory droplets---salt, lipids, and proteins---react to RH, and how this affects virus viability.

Salt ions are a ubiquitous component in physiological fluids, and when their concentration changes, the viability of enveloped and non-enveloped viruses exhibits different responses. Adding salt to a droplet medium has been shown to improve rather than reduce the viability of some non-enveloped viruses~\cite{Benbough1971a,Harper1963}, which is potentially related to the observations that screening of electrosatic interactions due to salt ions is in general beneficial to the stability of (often highly charged~\cite{Bozic2012,Siber2012}) non-enveloped viruses~\cite{Buzon2020,Comas2019,Garmann2014,Hagan2020,Lavelle2009,Perlmutter2014,Chen2018}. On the other hand, salt ions usually have a toxic effect on enveloped viruses, but the precise mechanisms remain unclear~\cite{Yang2012}. The way enveloped viruses acquire their lipid membrane varies from virus to virus~\cite{Garoff1998,Perlmutter2015,Welsch2007}, and in some cases the complexation of capsid protein and lipid membrane is driven by electrostatic interactions~\cite{Furukawa2020,Perlmutter2015}, which are influenced by changes in salt concentration. Salt ions also interact with lipid membranes and can cause structural and mechanical changes~\cite{Lee2008,Pabst2007,Vacha2009,Valley2011,Petrache2006}, potentially leading to inactivation of enveloped viruses~\cite{Benbough1969,Choi2015,Sooryanarain2015,Yang2012,Yang2012b}. The major role of ionic concentration for the assembly and stability of enveloped viruses thus seems related predominantly to the stability of lipid membranes and the interactions of various capsid proteins with the membranes.

It is also important to clarify the link between salt concentration, RH, and virus viability. As water evaporates from a respiratory droplet, salt ions become more concentrated, yet their concentration does not correlate with RH in a linear manner~\cite{Cohen1987,Lin2019,Yang2012}. Concentration of salt in the droplet residue depends on both droplet composition and RH: An initial concentration of $\sim 100$ mM NaCl can thus increase all the way up to $10$ M during evaporation~\cite{Lin2019}. Even higher concentrations can lead to an efflorescence transition, where salt crystallizes (at least partially) out of the aqueous part of the droplet (\Sec~\ref{ssec:efflorescence}). Yang \ea~\cite{Yang2012b, Yang2012} identified three regimes of RH for the viability of influenza A virus, closely related to the U-shaped viability curves (\Fig~\ref{fig:11}): {\em (i)} Close to $\rh=100\%$, salt concentration in the droplet stays at levels close to physiological conditions, and virus viability is thus well-preserved. {\em (ii)} Intermediate values of RH ($50\%$ to $\lesssim100\%$) involve concentrated and even supersaturated salt conditions that can be toxic to the virus, and viability consequently decreases with decreasing RH. {\em (iii)} In a very dry environment ($\rh\lesssim50\%$), salts can undergo efflorescence and crystallize out of the solution. The concentration of the remaining dissolved ions in the droplet residue is low, and consequently virus viability improves. If that is the case, the phenomenon of efflorescence turns out once again to be one of the culprits for the higher likelihood of droplet transmission at low RH, as it both reduces the size and sedimentation of droplets (\Sec~\ref{ssec:efflorescence}) and at the same time improves the viability of aerosolized viruses.

Respiratory droplets can differ to great extent in the amount of various proteins, biopolymers, and lipids they contain (Table~\ref{tab:composition}), which is important as studies have observed different responses of virus viability to changes in RH based on the origin and composition of droplets. Saliva has, for instance, been indicated to provide an important initial barrier to influenza A infection~\cite{White2009}, and both cell culture medium and artificial saliva have been shown to be more protective still~\cite{Zuo2014}. On the other hand, influenza virus has been observed to survive much longer on banknotes when in the presence of respiratory mucus~\cite{Thomas2008}, and extracellular material containing mucins has been shown to provide a (concentration-dependent) protective effect against RH-dependent decay of both influenza A and bacteriophage $\Phi$6~\cite{Kormuth2018}. At low RH, mucin has also been shown to protect viruses from damage by dehydration~\cite{Vejerano2018}. Peptones, lipids, and apolar amino acids also reduce virus losses, probably by protecting them against surface inactivation~\cite{Donaldson1976,Ehrlich1964,Hemmes1960,Trouwborst1972b,Trouwborst1974,Vejerano2018}, and protective effects have also been reported for polyhdroxy compounds~\cite{Benbough1971a,Schaffer1976}. Importantly, the protecting concentrations of these components can be related to the salt concentration in the droplet medium~\cite{Benbough1971a,Trouwborst1972b,Yang2012b}. The total composition of the droplet is the one determining how the droplet shrinks with time and whether or not it undergoes an efflorescence transition.

\subsection{Droplet pH}

Droplets can also vary in their pH~\cite{Effros2002,Freedman2018,Hunt2000,Wei2018}, which can change with RH and with that affect the viability of aerosolized viruses by changing the electrostatic properties or conformation of viral proteins~\cite{Bozic2018,Roshal2019,Taylor2002,Yang2012,Asor2020,Chevreuil2018,Cuellar2010}. The response of enveloped viruses to changes in RH further depends on whether the process of fusion requires low pH or not~\cite{Yang2012}. Viruses that require acidification before fusion (such as influenza virus and SARS-associated coronavirus) were found to be less stable at intermediate RH ($50\%$ to $90\%$) compared to higher and lower RH. Viruses that fuse at neutral pH (such as RSV) were found to be more stable at intermediate RH, and viruses that can fuse at both low and neutral pH (for example vaccinia virus and pigeon pox virus) were found to be insensitive to RH. pH also interacts with other variables: Stallknecht \ea~\cite{Stallknecht1990} observed a strong interactive effect between pH and salinity on the viability of influenza virus. The viability was highest at zero salt and high pH, high at high salt and low pH, but lowest at both high salt and high pH as well as low salt and low pH.

\subsection{Inactivation at the air-water interface}
\label{sssec:awi}

The fate of aerosolized viruses depends not only on the physico-chemical environment in the droplet but also on the precise location of the virus inside it, as droplets themselves can be internally heterogeneous~\cite{Vejerano2018} and viruses that adsorb at their air-water interface (AWI) can be inactivated to a great extent. The reasons are dehydration and unfavourable interactions with the AWI, including surface tension, shear stress, denaturation of proteins in contact with air, and conformational rearrangements driven by hydrophobicity~\cite{Casanova2010,Thompson1998,Thompson1999,Trouwborst1973,Zeng2017,Zuo2014}. The contribution of different mechanisms depends on RH, as droplet evaporation changes the AWI available for virus accumulation~\cite{Yang2012}. The extent to which a virus is attracted to the AWI is influenced both by the ionic strength and pH of the suspending medium as well as by the relative hydrophobicity, surface charge, and shape of the virus~\cite{Armanious2016,Okubo1995,Villa2020,Wan2002}. Both high ionic strength and large surface hydrophobicity (which differs from virus to virus~\cite{Heldt2017,Johnson2017}) create a high affinity for virus adsorption to the AWI~\cite{Thompson1999,Zuo2014}. Adsorption to the AWI is also influenced by changes in pH~\cite{Torres2016} (and therefore the net surface charge of the virus~\cite{Bozic2017,Bozic2018b}). Furthermore, enveloped viruses are more affected by surface inactivation at the AWI than non-enveloped ones~\cite{Donaldson1976}, either because of differences in their affinity for the adsorption to the AWI or because they are not affected by interfacial forces in the same manner as non-enveloped viruses~\cite{Mitev2002,Thompson1999}. Surface-active compounds in the droplet, such as various proteins, amino acids, and surfactants, can accumulate at the AWI~\cite{Bzdek2017} and in doing so prevent the aerosolized viruses from reaching the AWI and being inactivated~\cite{Thompson1999}, showing once more the importance of droplet composition for the survivability of enclosed viruses.

\subsection{Experimental factors}

The observations relating virus viability to changes in RH are further complicated by the interplay between temperature and humidity~\cite{Hermann2007,Moe1983,Tang2009,Zhao2012} and the varying conditions under which the experiments are performed (\eg, droplet composition and aerosol ageing; \Fig~\ref{fig:11})~\cite{Pica2012,Sattar1987,Tang2009,Yang2012,Yang2012b}. Often, droplet composition is simplified in both experimental and theoretical models and comprises salt ions, proteins, and surfactants to different extents~\cite{Pica2012,Sattar1987,Tang2009}. For instance, salt ions are not always included in model respiratory droplets, and when they are, their concentrations range anywhere from $30$ to $300$ mM~\cite{Marr2019,Vejerano2018,Yang2012b}. The exact nature and concentration of salt ions or proteins in model respiratory droplets could thus prove critical in determining the RH-dependent response of virus viability upon aerosolization. Different artificial means of producing virus aerosols may also not be comparable to the natural release of viruses in saliva or respiratory mucus~\cite{Vejerano2018,Zuo2014}, which can act as an organic barrier against environmental extremes~\cite{Tang2009}. The composition of both saliva and respiratory mucus is a complex mixture containing different electrolytes, proteins, and surfactants, with each of the components exceeding the mass of the virus by several orders of magnitude~\cite{Lin2019,Vejerano2018,Zuo2014} and potentially influencing virus survival~\cite{Humphrey2001,Malamud2011,Zanin2016}.

\section{Conclusions}
\label{sec:concl}

Seasonal periodicity of respiratory infections in humans is driven by complex mechanisms, ranging from environmental to social. Mounting evidence suggests that a critical player in the observed seasonality is the RH of air inside buildings. Low indoor RH, as experienced during winter months or inside airplanes, directly or indirectly influences several mechanisms that increase the transmission of respiratory diseases. In this review, we attempted to summarize and elucidate those known physical mechanisms of droplet and airborne transmission that are influenced by RH.

RH starts to play a role the moment respiratory droplets are exhaled into the air. Dry air accelerates droplet evaporation while at the same time dehydrates them more, so that the size of droplet residues after evaporation has stopped is smaller than in more humid air. Both effects cause the droplets to settle to the ground more slowly and remain in the air longer at low RH, thereby leaving more potential pathogens suspended in the air. RH furthermore governs the extent to which viruses carried by the droplets survive. This relationship is complex and depends both on the droplet composition as well as on the structure of the virus, where enveloped viruses tend in general to be more vulnerable to environmental changes. Quite remarkably, the viability of several viruses improves when RH is lowered below $50\%$, which is also one of the explanations for the strong seasonality of the influenza virus. An important yet less noticed phenomenon that accompanies evaporation of respiratory droplets is efflorescence, which, as evidence suggests, occurs at least to some extent in respiratory droplets. It accentuates the effects of RH by rapidly changing droplet size, consequently affecting the detailed nature of droplet composition.

Indoor RH during winter in temperate climates is typically around $10\%$ to $20\%$ lower than during summer, and an important reason for this are heating and poor ventilation of indoor spaces~\cite{Sundell2011,Hobday2013,Bentayeb2015}. This review shows that maintaining indoor RH at $\approx50\%$ is not only the most comfortable level for humans but perhaps also a good target value to aid in preventing the spread of infectious diseases. A better understanding of where and how it affects the behavior of droplets and any pathogens contained in them would allow us to exploit this knowledge to control indoor RH in such a way as to minimize the spread of droplet- and aerosol-borne disease.

\begin{acknowledgements}
We thank Roland R.\ Netz, Rudolf Podgornik, and Luca Tubiana for fruitful discussions and comments.
We acknowledge funding from the Slovenian Research Agency ARRS (Research Core Funding No.\ P1-0055 and Research Grant No.\ J1-1701).
\end{acknowledgements}

\bibliographystyle{spphys}
\bibliography{references}

\end{document}